\documentclass[preprint,12pt]{article}

\usepackage{amssymb}

\usepackage{amsmath}

\usepackage[table]{xcolor}
\usepackage{graphicx}
\usepackage{tabularx}
\usepackage{makecell} % 
\usepackage{xcolor}
\usepackage{float}
\usepackage{subcaption}
\usepackage{colortbl}
\usepackage{textcomp}
\usepackage{multirow}  % in preamble
\usepackage{array}
\renewcommand{\arraystretch}{1.2}
\usepackage{booktabs}
\usepackage{textgreek}
\usepackage{hyperref}
\usepackage[utf8]{inputenc}
\usepackage{bm}
\usepackage{rotating}
\usepackage{rotating}
\usepackage{multirow}
\usepackage{caption}
\usepackage{float}
\usepackage{booktabs}
\usepackage{array}
\usepackage{authblk}
\usepackage[labelfont=bf]{caption}
\usepackage[justification=justified]{caption}
\usepackage[a4paper, margin=1in]{geometry}
\usepackage{fancyhdr}
\begin{document}

% \title{CORE: A Cell-Level Coarse-to-Fine Image Registration Engine for Multi-stain Image Alignment}
\title{CORE: A Coarse-to-Fine Registration Engine for Multi-stain Whole-Slide Images}
% \title{CORE: Cell-aware Coarse-to-Fine Registration for Multi-stain Whole-Slide Images}
% \titlerunning{COarse-to-fine Registration Engine}

\author[1,2]{{Esha Sadia Nasir}}
% \ead{esha.nasir@warwick.ac.uk}
\author[1,2]{{Behnaz Elhaminia}}
\author[1]{{Mark Eastwood}}
\author[3]{Catherine King}
\author[4]{Owen Cain}
\author[3,5]{Lorraine Harper}
\author[3]{Paul Moss}
\author[3,4]{Dimitrios Chanouzas}
\author[2,7]{{David Snead}}
\author[2,7]{{Nasir Rajpoot}}
\author[2]{{Adam Shephard}}
\author[1,2]{Shan E Ahmed Raza}
% \ead{shan.raza@warwick.ac.uk}

\affil[1]{VISION Lab, Department of Computer Science, University of Warwick, UK.}
\affil[2]{Tissue Image Analytics (TIA) Centre, Department of Computer Science, University of Warwick, UK.}
\affil[3]{Department of Immunology and Immunotherapy, College of Medicine and Health, University of Birmingham, UK.}
\affil[4]{Renal Unit, Queen Elizabeth Hospital Birmingham, University Hospitals Birmingham NHS Foundation Trust, UK.}
\affil[5]{Department of Cellular Pathology, Queen Elizabeth Hospital Birmingham, UK.}
\affil[6]{Department of Applied Health Sciences, College of Medicine and Health, University of Birmingham, UK}
\affil[7]{Histofy Ltd, Coventry, UK.}
\date{}
\maketitle
\begin{abstract}
% Problem (WSI registration importance)
% Challenge (variability, deformation)
% Gap (lack of generalization)
% Contribution (CORE)
% Method summary (coarse-to-fine pipeline)
% Evaluation + results

Accurate registration of multi-stained whole-slide images (WSIs) enables the integration of complementary morphological and molecular information, improving both clinical diagnostic and prognostic analysis. It also enables efficient transfer of annotations between consecutive or re-stained slides, significantly reducing annotation time and cost, as well as three-dimensional tissue reconstruction. Despite its importance, WSI registration remains challenging due to variations in slide preparation across stains and complex tissue deformations. Existing methods are often tailored to specific staining modalities and fail to generalize across diverse settings, highlighting the need for a robust and generalizable cell-level multi-stain registration method. In this work, we introduce CORE, a novel coarse-to-fine framework for accurate cell-centric registration across \textbf{six} multimodal WSI datasets, encompassing \textbf{30} distinct staining types, including Hematoxylin \& Eosin (H\&E), periodic acid–Schiff (PAS), multiplex immunohistochemistry (mIHC), multiplex immunofluorescence (mIF) and Cyclic immunofluorescence (Cyc-IF). CORE first performs coarse global registration by extracting tissue masks through prompt-based segmentation to remove artifacts and non-tissue regions, followed by rigid alignment using tissue morphology and a pre-trained feature extractor. The resulting alignment is refined using a shape-aware point-set registration model applied to automatically detected nuclei centroids, enabling fine-grained rigid alignment. Finally, Coherent Point Drift (CPD) is used to estimate a non-linear displacement field for non-rigid cellular alignment. Using nuclei correspondences throughout the pipeline, CORE achieves accurate and robust cell-level alignment across modalities.

% We evaluate CORE on six datasets, including three public benchmarks and two private datasets. 
Experimental results show that CORE consistently outperforms state-of-the-art methods in accuracy, robustness, and generalization across both bright-field and immunofluorescence WSIs.
% To our knowledge, this is the only study incorporating both tissue and cell-level registration for WSI alignment across such a large and diverse set of datasets and stain types. 

\textbf{Keywords:} 
Registration, multi-stain, nuclei, alignment, rigid, non-rigid, deformation 

\end{abstract}

\newcolumntype{P}[1]{>{\RaggedRight\footnotesize}p{#1}}

\section{Introduction}
WSI registration is a fundamental task in computational pathology, enabling the spatial alignment of corresponding tissue sections across different stains, imaging modalities, and acquisition settings. By establishing precise correspondence between tissue structures, WSI registration allows integration of complementary information from multiple sources, thus supporting a broad spectrum of downstream applications including biomarker discovery, tumor microenvironment (TME) analysis \cite{trahearn2017hyper}, annotations transfer \cite{su2022deep} \ref{fig:annotation_transfer} \footnote{Supplementary Figure~\ref{fig:annotation_transfer} illustrates glomeruli annotations transferred from bright-field to florescence re-stained sections using the proposed CORE registration framework.}, multimodal image fusion, and three-dimensional tissue reconstruction \cite{elhaminia_traditional_2025}. For example, multimodal registration between H\&E and mIF images enables direct integration of morphological and molecular information at cellular resolution, supporting downstream prognostic and biomarker analyses~\cite{lin_high-plex_2023}. Registration also facilitates efficient transfer of annotations between corresponding tissue sections, substantially reducing the time and cost associated with manual labeling. 
% Phosphohistone H3 (PHH3) stained WSIs can be registered to H\&E slides to support mitotic figure detection and classification \cite{nasir_mitodetect_2025}.
For instance, in our subsequent work, Ki67-to-H\&E registration facilitated the transfer of mitotic annotations, enabling accurate mitosis detection and classification using only routinely acquired H\&E slides~\cite{ali_mitonet_2026}. This leads to cost savings since H\&E staining is relatively cheap and more widely available. WSI registration has also been applied to serial tissue reconstruction, enabling volumetric recovery of tissue architecture from sequential histological sections \cite{gatenbee_source_2023, feuerstein2011reconstruction, kiemen2022coda}.

\begin{figure}[H]
\centering
\begin{minipage}{\textwidth}
    \centering
    \includegraphics[width=\textwidth]{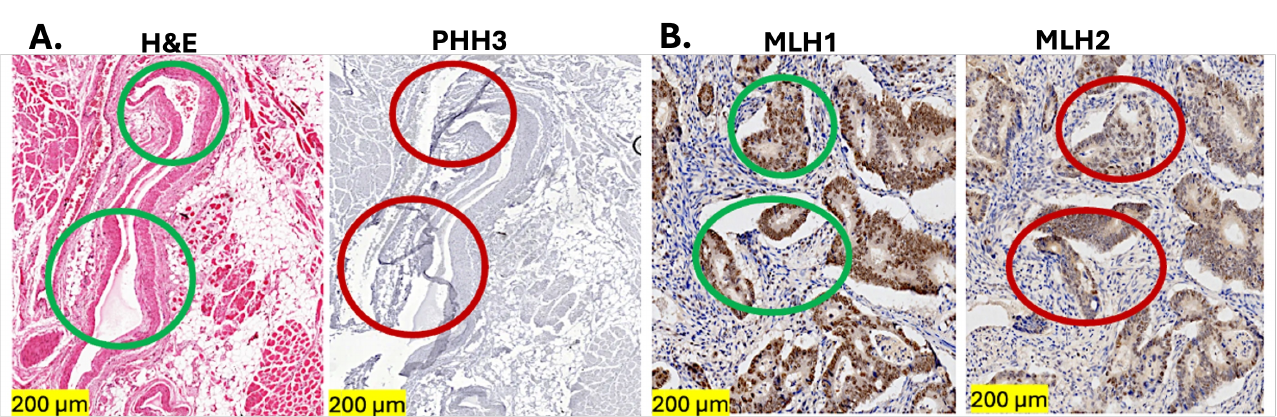}
    \captionsetup{width=\textwidth}
    \caption{Left block shows re-stained sections from the HyReCo dataset. (A) H\&E-stained region extracted from a WSI. Re-stained PHH3 section; red circles indicate staining artifacts, indicating missing tissue and tissue fold regions visible only at high-resolution. Corresponding regions are highlighted with green circles in the H\&E image. (B) The block shows two consecutive sections from the Multi-IHC CRC dataset. MLH1-stained region extracted from a WSI. MLH2-stained section; red circles indicate morphological variations in tissue structure, as staining is applied to different sections, unlike re-stained samples where the same section is used. Corresponding regions are highlighted with green circles in the MLH1 image to illustrate sectioning variation.}
    \label{fig:cons_re-stained}
\end{minipage}
\end{figure}

Despite its broad applicability, robust multimodal WSI registration remains a challenging problem. 
It is commonly performed between either re-stained sections (derived from the same tissue) or consecutive sections originating from adjacent physical slices, each presenting distinct alignment challenges~\cite{elhaminia_traditional_2025}. 
Re-stained slides originate from the same tissue section and therefore generally permit more accurate alignment. In contrast, consecutive sections are particularly difficult to align because corresponding cellular structures may partially disappear, deform, or shift between neighboring slices~\cite{lotz_comparison_2023}. Both settings are further complicated by non-linear tissue deformations introduced during tissue preparation and digitization, including tearing, folds, stretching, staining artifacts, and partial tissue loss~\cite{elhaminia_traditional_2025}. These challenges are amplified in multimodal WSIs because different stains such as H\&E, mIHC, PAS,  and mIF emphasize distinct biological structures and tissue components. Consequently, structures that are highly visible in one modality may appear weak, distorted, or entirely absent in another, making reliable cross-stain correspondence estimation difficult. Figure~\ref{fig:cons_re-stained} shows re-stained and consecutive staining variation in different stains and challenges in terms of registration for each section. In addition, the gigapixel scale of WSIs introduces substantial computational and memory constraints that limit the applicability of conventional registration techniques.

Recent advances in WSI registration have introduced a variety of rigid, deformable, feature-based, and deep learning-driven registration strategies. While these methods have shown promising performance for specific stain combinations and controlled experimental settings, several important limitations remain unresolved. First, many existing approaches are optimized for a limited set of modalities, most commonly H\&E and IHC \cite{lotz_robust_2019}, limiting generalization across heterogeneous stain domains. Second, most frameworks predominantly optimize alignment at the coarse (tissue or image) level without explicitly modeling cellular correspondence at high-resolution, thereby failing to capture local deformations. Third, existing evaluations are often restricted to individual datasets or narrowly defined registration settings, making it difficult to assess robustness across diverse tissue types, staining protocols, and sectioning conditions.

To address these limitations, we propose CORE, a hierarchical registration framework that progressively refines alignment from global tissue morphology to fine-grained nuclei correspondence. Unlike prior approaches that rely on image-level similarity or stain-specific appearance features, CORE integrates tissue-level structural alignment with cell-centric point-set registration to achieve robust multimodal registration across heterogeneous staining protocols and tissue sectioning conditions. The framework combines morphology-guided coarse alignment, accelerated feature refinement, and nuclei-driven deformable registration within a unified coarse-to-fine model.
\subsection{Contributions}

The principal contributions of this work are summarized as follows:

\begin{itemize}

\item \textbf{High resolution multimodal WSI registration framework:}  We propose CORE, a coarse-to-fine multimodal WSI registration framework that combines global tissue-level alignment with local fine-grained nuclei-level refinement for robust multi-stain histopathology registration.

\item \textbf{Cell-centric shape-aware registration formulation:}  We develop a shape-aware nuclei point-set registration framework that incorporates both spatial and morphological information for accurate cell-centric alignment under cross-section and cross-stain variability.

\item \textbf{Evaluation across heterogeneous stain modalities: } CORE demonstrates robust multimodal registration across \textbf{30} distinct stain types, \textbf{six} different datasets spanning H\&E, PAS, mIHC, mIF and Cyc-IF modalities, including both re-stained and consecutive tissue sections. To the best of our knowledge, CORE is the first registration framework to be comprehensively evaluated on such a large and diverse collection of WSIs spanning multiple staining technologies and tissue preparation protocols. Details of all stain types are in  Table (\ref{tab:stains}).

\item \textbf{Efficient gigapixel-scale registration:} We combine morphology-guided alignment with efficient feature refinement to significantly accelerate coarse registration, reducing computational overhead while improving alignment accuracy across WSIs.

\item \textbf{Visualization tool and open-source codes:} We provide an interactive, real-time visualization framework for exploring both rigid and non-rigid deformations across entire WSIs at full resolution, enabling intuitive inspection and validation of registration quality along with the implementation code. To the best of our knowledge, no publicly available WSI registration framework offers real-time visualization of deformation fields directly on full-resolution WSIs. A working demonstration of \textbf{CORE} is available through TiaViz\footnote{\url{https://tiademos.dcs.warwick.ac.uk/bokeh_app?demo=WSIReg}}, with additional details provided in Appendix~\ref{app:tiaviz}.
\end{itemize}

The remainder of this paper is organized as follows. Section 2 reviews related WSI registration methods. Section 3 presents the proposed CORE framework. Section 4 describes the datasets and evaluation metrics. Section 5 reports experimental results followed by discussion and conclusions in Section 6. Supplementary information provides additional details, quantitative tables and figures of this study.

\section{Related Work}

WSI registration remains a fundamental challenge in computational pathology due to large image resolution, heterogeneous tissue morphology, staining variability, and tissue deformations introduced during slide preparation. Existing WSI registration methods can broadly be categorized into intensity-based, feature-based, and deep learning based approaches.
Feature-based methods estimate sparse correspondences using handcrafted or learned descriptors, while intensity-based approaches optimize image similarity measures directly from pixel intensities. Common similarity metrics include cross-correlation, mutual information, and normalized gradient fields. In practice, most frameworks combine coarse global alignment with subsequent deformable refinement.
A detailed review of the existing WSI registration methods, datasets, tools and their limitations has been presented in Elhaminia et al.~\cite{elhaminia_traditional_2025}. Representative multimodal WSI registration frameworks include HistokatFusion~\cite{lotz_comparison_2023}, VALIS ~\cite{gatenbee_virtual_2023}, DeeperHistReg ~\cite{wodzinski_regwsi_2024}, and DFBR ~\cite{awan_deep_2023}. Among these methods, HistokatFusion primarily rely on iterative optimization strategies, whereas DFBR, DeeperHistReg and VALIS incorporate deep feature extraction for correspondence estimation.

Early WSI registration methods relied primarily on iterative intensity-based optimization. Lotz et al.~\cite{lotz_patch-based_2016}, combined low-resolution global alignment with patch-level refinement to estimate local tissue deformations. Although effective for localized alignment, the framework struggled to model spatial deformations consistently across entire WSIs. 
This limitation motivated the development of HistokatFusion~\cite{lotz_comparison_2023}, a multi-stage model combining affine initialization, Gauss–Newton optimization, and B-spline deformable refinement. The framework demonstrated strong performance on the ANHIR~\cite{borovec_anhir_2020} and ACROBAT benchmarks~\cite{borovec_anhir_2020}. Nevertheless, the method was developed primarily for H\&E and IHC registration and is therefore less suited to highly heterogeneous modalities such as mIF. In addition, the implementation is not currently publicly available.

% VALIS

Similarly, later methods focussed on the development of hybrid frameworks that combined handcrafted descriptors with learned feature representations.  
Gatenbee et al.~\cite{gatenbee_virtual_2023} introduced VALIS, a multimodal registration framework integrating preprocessing, rigid alignment, and deformable refinement for sequential H\&E, mIHC, and mIF images. The framework combines handcrafted and learned feature representations to improve robustness across staining modalities. However, its iterative optimization strategy for nonrigid registration remains computationally demanding for gigapixel WSIs and fails in the presence of severe staining artifacts or missing tissue regions.
Additionally some studies have also explored structural and morphological features for WSI registration. For example, Sarkar et al.~\cite{sarkar_robust_2014} employed tissue boundary features and RANSAC-based alignment for adjacent tissue sections, while Ozturk and Akdemir~\cite{ozturk_comparison_2018,ozturk_effective_2018} investigated classical descriptors including HOG, MSER, SIFT, and LBP for histopathological feature extraction. These studies highlighted the importance of robust structural representations and efficient preprocessing for scalable WSI analysis.

Recent advances in computational resources and the availability of large-scale datasets have accelerated the development of deep learning based WSI registration frameworks. A notable example is DeeperHistReg, proposed by Wodzinski et al.~\cite{wodzinski_regwsi_2024}, which combines deep feature extraction using SuperPoint~\cite{ge_unsupervised_2022} with feature matching via SuperGlue~\cite{sarlin_superglue_2020}, followed by intensity-based non-rigid refinement. The framework employs a pyramidal multi-resolution strategy to address GPU memory constraints while improving registration performance. However, the method was primarily developed for H\&E and mIHC registration and demonstrates limited generalization to more challenging modalities such as mIF.  Similarly, Awan et al.~\cite{nixon_deep_2018} further explored deep representation learning for multimodal WSI registration. They introduced an autoencoder-based framework capable of learning feature representations directly from paired images and estimating transformations through gradient-based optimization. This approach later evolved into DFBR~\cite{awan_deep_2023}, a multi-stage framework combining preprocessing, rigid alignment, and deformable refinement. For this, tissue masks were first generated for coarse transformation estimation, followed by rigid alignment using multi-scale CNN (VGG) feature extraction and feature matching. Residual alignment errors were further corrected using phase correlation before applying the deformable registration strategy proposed by Lotz et al.~\cite{lotz_robust_2019}.

Current research has increasingly focused on incorporating biologically meaningful structural information into multimodal registration.
Instead of relying solely on image intensities or handcrafted descriptors, new methods now leverage segmentation-derived tissue structures and nuclei-level representations to improve robustness across staining modalities. Ge et al.~\cite{ge_unsupervised_2022} proposed a structural feature-guided CNN capable of capturing both low-resolution global structure and high-resolution tissue details for multi-stain registration. Similarly, Mahapatra et al.~\cite{mahapatra_registration_2020} employed segmentation maps generated by a pre-trained U-Net to guide deformable alignment. These approaches marked a transition from intensity-driven registration toward structurally informed alignment. Huang et al. \cite{wei_unsupervised_2025} employed morphological descriptors, including area and eccentricity of segmented red blood cells, while Cooper et al.~\cite{antonacopoulos_unsupervised_2025} incorporated broader feature sets including centroid, area, eccentricity, and axis orientation from segmented H\&E and CD3 stained images. This shift is particularly evident in nuclei-based registration approaches. Jeyasangar et al.~\cite{yap_nuclei-location_2024} proposed a nuclei-driven WSI registration framework that detects nuclei using Hover-Net, extracts nuclei point sets from image tiles, and aligns them using a Gaussian Mixture Model (GMM) combined with local linear embedding to preserve spatial organization. Their approach demonstrated improved performance on the HyReCo dataset~\cite{van_der_laak_hyreco_2021} and generalization across staining variations. However, the evaluation focused primarily on patch level registration and a limited subset of stains. Similarly, Jiang et al.~\cite{jiang_multimodal_2024} proposed a multimodal registration framework for aligning H\&E and mIF images at the cellular level. Their method treated segmented nuclei as point sets and employed CPD followed by graph matching refinement to achieve accurate cellular alignment. Evaluated on ovarian cancer tissue microarrays, the framework enabled integration of multimodal cell-level information. 
Despite these advances, existing nuclei-aware frameworks remain limited either to patch-level registration, restricted stain combinations, or tissue microarray settings, and do not provide a unified hierarchical framework for full-resolution multimodal WSI registration.

For optimization, bucak and uslan~\cite{bucak_sequence_2011} surveyed the application of stochastic optimization (SO) to biological sequence alignment, an NP-complete problem. They found that as datasets grow, traditional Dynamic Programming (DP) becomes computationally prohibitive due to exponential growth in time and memory requirements. The paper further highlights various iterative heuristics including simulated annealing, genetic algorithms, and swarm intelligence (ACO and PSO) that provide faster, near-optimal solutions. Specifically, methods like ant colony optimization (ACO) are noted for their robustness and ability to escape local optima in complex computational biology tasks. Followed by a hybrid stitching technique for cytopathologic examination that automates the creation of 2D and 3D panoramic images. Using a modified Iterative Closest Point (ICP) algorithm, they extracted feature points from 2D images via SURF and projected them into a 3D space to calculate a $4 \times 4$ transformation matrix. This approach addresses light microscopy limitations, such as narrow fields of view and limited depth of field, while reducing artifacts like seams and outliers. The method demonstrated superior efficiency, achieving execution times as low as 9.84 seconds with high quantitative accuracy~\cite{dogan_hybrid_2021}.

Collectively, these studies illustrate the evolution of WSI registration from classical iterative optimization towards hybrid, deep learning-driven, and structure-aware frameworks. Despite substantial advances, several important challenges remain unresolved. Existing methods are often evaluated  either at coarse or fine resolution levels and are often optimized for specific stain combinations such as H\&E-to-mIHC~\cite{awan_deep_2023, lotz_comparison_2023, ali_swift-reg_2026} or H\&E-to-DAPI~\cite{gatenbee_virtual_2023, doyle_whole-slide_2023}. Furthermore, many approaches remain sensitive to staining variability, heterogeneous tissue morphology, and severe local deformations due to reliance on stain-specific appearance cues or computationally intensive optimization procedures.
 
To address these limitations, we propose CORE, a hierarchical multimodal WSI registration framework that progressively refines alignment from tissue-level morphology to nuclei-level correspondence. The framework integrates coarse structural alignment, feature-guided refinement, and cell-centric deformable registration within a unified coarse-to-fine formulation capable of handling both re-stained and consecutive tissue sections across \textbf{six} datasets encompassing \textbf{30} distinct staining modalities.

\section{Method}

We propose CORE, a coarse-to-fine framework for multimodal WSI registration that progressively refines alignment from global tissue correspondence to local cellular-level alignment. The framework is designed to address stain-induced morphological differences, structural heterogeneity between serial sections, and complex non-linear tissue deformations commonly encountered in multimodal WSIs, including H\&E, PAS, mIHC, CyC-IF and mIF images.

Given a source WSI $S$ and a target WSI $T$ the objective is to estimate a spatial transformation $\mathcal{T}$ that aligns $S$ to $T$ despite differences in staining, tissue preparation, and local tissue morphology. Direct end-to-end registration at full WSI resolution is computationally prohibitive and often unstable due to the gigapixel scale of the images, substantial inter-stain variability, and limited cross-modal correspondence. To address these challenges, we formulate WSI registration as a  coarse-to-fine alignment problem that progressively refines alignment through sequential global and local registration stages. The main components of CORE includes:

\begin{enumerate}

    \item Preprocessing (processes tissue for improved stain appearance and tissue-mask extraction)
    \item Coarse Registration (Tissue-Level Alignment)
    \item Fine-Grained Registration (Cell-level Alignment)
    \item Registration Visualization (TiaViz)
\end{enumerate}

The overall transformation is defined in Eq. (\ref{eq:overall})

\begin{equation}
\mathcal{T} =
\mathcal{T}_{\text{CPD}}
\circ
\mathcal{T}_{\text{shape}}
\circ
\mathcal{T}_{\text{deform}}
\circ
\mathcal{T}_{\text{rigid}}
\label{eq:overall}
\end{equation}
where:

\begin{itemize}
 \item    $\mathcal{T}_{\text{rigid}}$ denotes the coarse rigid transformation,
    \item $\mathcal{T}_{\text{deform}}$ represents coarse deformable alignment,
    \item $\mathcal{T}_{\text{shape}}$ corresponds to nuclei-level shape-aware refinement, and
    \item $\mathcal{T}_{\text{CPD}}$ models local non-linear cellular deformation.
\end{itemize}

This hierarchical decomposition separates large-scale geometric correction from fine cellular refinement, enabling robust multimodal registration while maintaining computational efficiency. An overview of the proposed framework is shown in Figure~\ref{fig:overall}.

\begin{center}
\captionsetup{width=\textwidth}

    \includegraphics[width=1\textwidth]{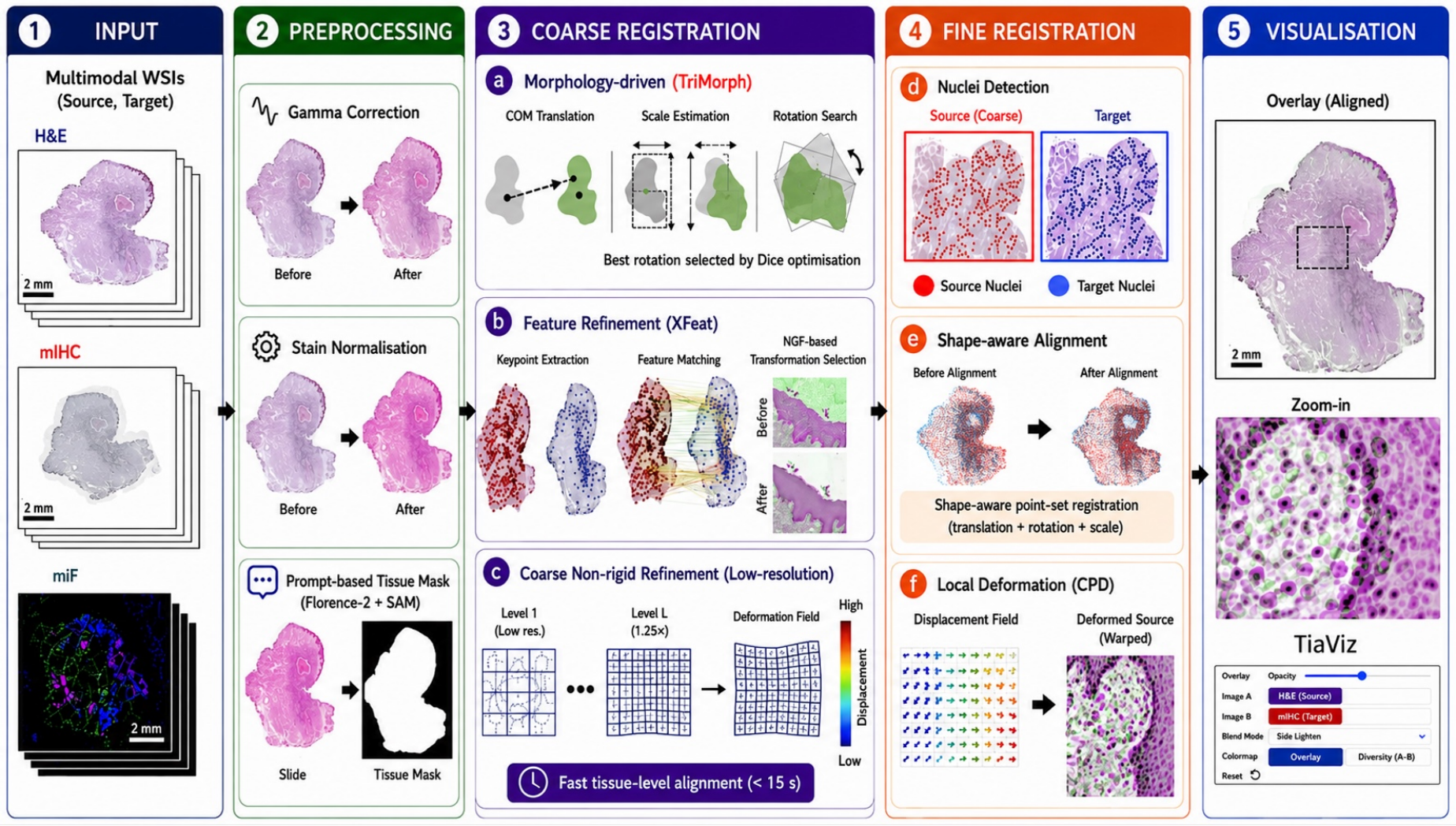}
 \captionof{figure}{
    Overview of the CORE multimodal WSI registration pipeline. 
    The framework aligns H\&E, mIHC, PAS, Cyc-IF and mIF WSIs through a coarse-to-fine strategy. 
    \textbf{(1)} Multi-stained WSI pairs from different staining protocols are provided as input. 
    \textbf{(2)} Preprocessing applies stain normalization, gamma correction, and prompt-based tissue masking using Florence-SAM. 
    \textbf{(3)} Coarse tissue-level registration proceeds in three steps: 
    \textbf{(a)} morphology-driven alignment via TriMorph, 
    \textbf{(b)} feature refinement using XFeat with NGF-based final transformation selection, and 
    \textbf{(c)} low-resolution non-rigid deformation. 
    \textbf{(4)} Fine cell-level registration involves: 
    \textbf{(d)} sparse nuclei detection in both images, 
    \textbf{(e)} shape-aware point-set alignment, and 
    \textbf{(f)} local deformation using CPD. 
    \textbf{(5)} The registered overlay is visualized along with visualization from TiaViz.}
    \label{fig:overall}
\end{center}

The first step of this pipeline is preprocessing, which standardizes the input representations to ensure that subsequent alignment operates on  consistent inputs across modalities.
\subsection{Preprocessing}

To enable reliable extraction of tissue morphology across diverse staining modalities, each WSI is first downsampled and preprocessed using gamma correction to improve tissue contrast and suppress background artifacts. Given an input image $I \in \mathbb{R}^{H \times W \times 3}$, gamma correction is applied to enhance weakly stained tissue regions and improve the visibility of morphological structures, as defined in Eq.~(\ref{eq:gamma}):

\begin{equation}
I' = I^\gamma
\label{eq:gamma}
\end{equation}

where $\gamma \in [1.0,1.2]$ for H\&E images and $\gamma \in [0.4,0.8]$ for IHC images.

For mIF WSIs, the DAPI channel is used as the primary structural reference due to its consistent nuclear representation across samples. To further reduce inter-slide color variability, Macenko stain normalization is applied using standard parameter settings ($I_0 = 240$, $\alpha = 1$, $\beta = 0.15$) and predefined stain vectors for H\&E, given in Eq.~(\ref{eq:hematoxylin}) and Eq.~(\ref{eq:eosin}), respectively:

\begin{equation}
H = [0.650, 0.704, 0.286]
\label{eq:hematoxylin}
\end{equation}

\begin{equation}
E = [0.072, 0.990, 0.105]
\label{eq:eosin}
\end{equation}

\paragraph{Prompt-based Tissue Mask Extraction:}

Following preprocessing, each WSI is processed to obtain a tissue foreground mask that defines the spatial domain for registration. This step is important as it restricts alignment to tissue regions. It also excludes the background slide content thus ignoring artifacts outside tissue regions i.e., dust, pen markings, slide background, borders and small debris. We use a prompt-guided segmentation strategy that combines the Florence-2 vision-language model~\cite{xiao_florence-2_2023} with the Segment Anything Model (SAM)~\cite{kirillov_segment_2023}. Florence-2 generates semantic prompts that guide SAM to produce binary tissue masks. A cascaded prompt set (\textit{tissue}, \textit{stain}, \textit{histology}, and \textit{cell}) is applied sequentially. It extracts segmented tissue excluding background artifacts across staining modalities. This prompt-driven formulation enables consistent tissue masking in diverse datasets and staining modalities. In rare cases (2–3\%), typically due to low contrast or severe staining degradation, the prompt-based method fails to produce a valid mask. In such cases, CORE automatically applies a U-Net model trained on the ACROBAT dataset~\cite{weitz_acrobat_2024} to generate tissue masks, thereby improving robustness and ensuring reliable registration performance. Morphological opening and closing with elliptical kernels is applied to remove residual noise regions, smooth tissue boundaries, and fill small discontinuities. Additionally, connected-component analysis is performed to retain only the largest tissue component. This eliminates small, detached artifacts such as control tissue or fragmented background regions, yielding a clean tissue foreground mask. Figure~\ref{fig:tissue_masks} shows tissue masks extracted by our prompt-guided tissue mask segmentation method.

\subsection{Coarse Registration (Tissue-Level Alignment)}

The objective of coarse registration is to estimate robust global alignment between source and target WSIs at low resolution ($0.625\times$--$1.25\times$). This stage resolves large-scale translation, rotation, scaling, and global tissue deformation prior to high-resolution cellular alignment. The coarse registration stage consists of:
\begin{enumerate}

    \item TriMorph: Morphology-driven Rigid Registration.
    \item Feature-based Refinement Using XFeat, and
    \item Coarse Non-Rigid Registration.
    
\end{enumerate}

\begin{figure}[H]
    \centering
    \includegraphics[width=\textwidth]{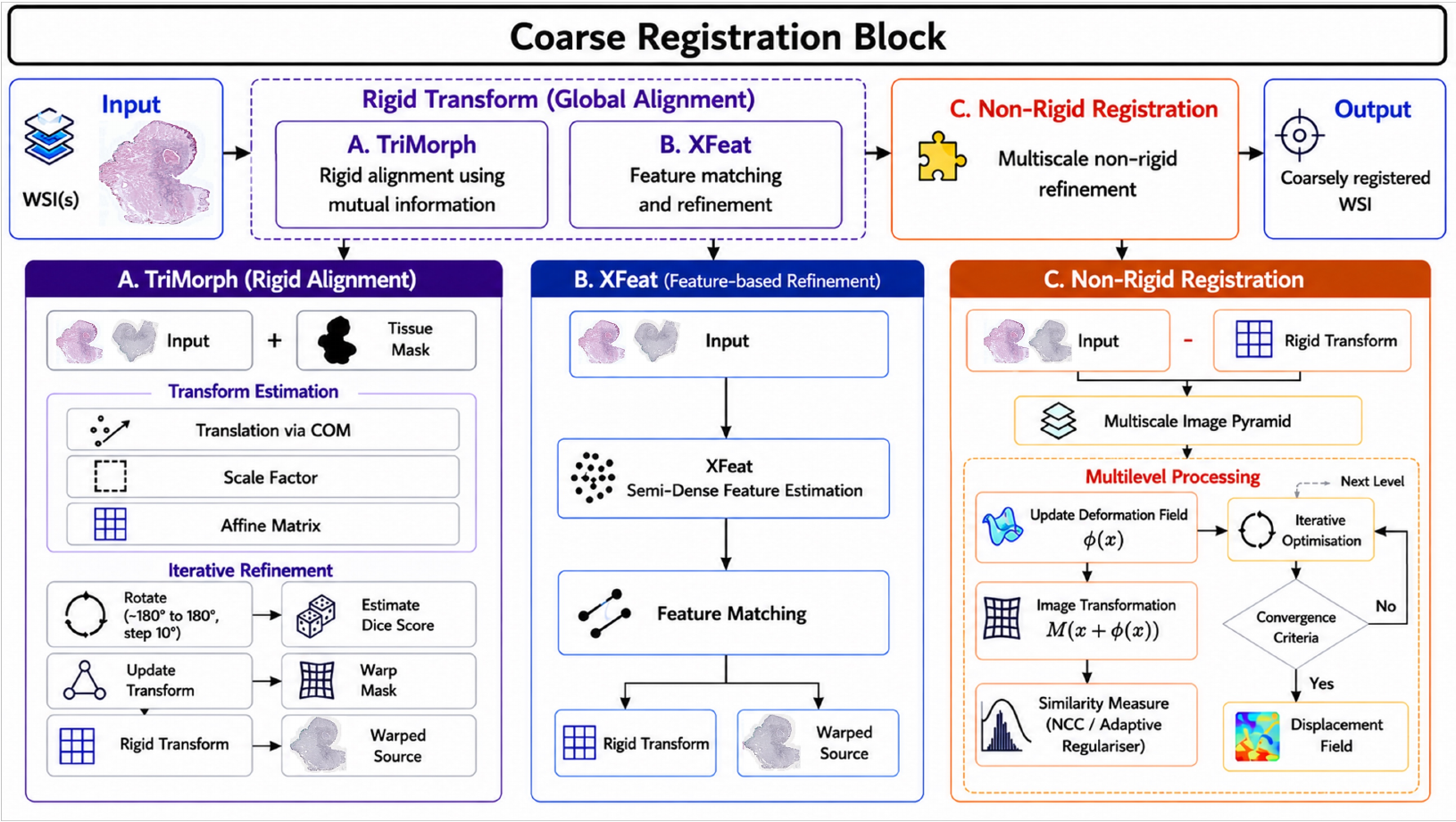}
    \captionsetup{width=\textwidth}
    \caption{Overview of the CORE \textbf{coarse registration }block for WSI alignment. (A) \textbf{TriMorph} estimates global rigid alignment by computing translation, scale, and rotation from tissue morphology. (B) \textbf{XFeat} performs feature-based refinement by extracting semi-dense correspondences and computing robust feature matches for rigid transform correction. (C) \textbf{Non-rigid} registration applies multiscale deformable optimization, where a displacement field is iteratively updated under similarity constraints (e.g., NCC) and adaptive regularization until convergence.}
    \label{fig:coarse_block}
\end{figure}

Figure~\ref{fig:coarse_block} shows a schematic diagram of CORE coarse registration block including coarse rigid and non rigid registration.

\subsubsection{TriMorph: Morphology-driven Rigid Registration}
We introduce \textbf{TriMorph}, as a tissue morphology-driven rigid alignment module that estimates translation, scaling, and rotation directly from binary tissue masks. Unlike intensity-based methods, TriMorph relies exclusively on tissue geometry, improving robustness under substantial cross-stain appearance variation. Let

\begin{equation}
M_S : \mathbb{R}^2 \rightarrow \mathbb{R}
\label{eq:source_mask}
\end{equation}

and

\begin{equation}
M_T : \mathbb{R}^2 \rightarrow \mathbb{R}
\label{eq:target_mask}
\end{equation}

Eqs. {(\ref{eq:source_mask}) and (\ref{eq:target_mask})} denote the source and target tissue masks, respectively. The rigid transformation is estimated through a sequential estimation of translation, scale normalization, and rotation search. First, centroid alignment is performed using center-of-mass (CoM) based translation. A scaling factor is then estimated from the relative spatial extent of the aligned masks. Finally, rotation shift is calculated over a discrete candidate set with angular step size $\theta = 10^\circ$. Each candidate transformation is evaluated using the Dice overlap coefficient defined in Eq.~(\ref{eq:dice}):

\begin{equation}
\text{Dice}(M_S^\mathcal{T},M_T)
=
\frac{
2|M_S^\mathcal{T} \cap M_T|
}{
|M_S^\mathcal{T}| + |M_T|
}
\label{eq:dice}
\end{equation}

The optimal rigid transformation is obtained by maximizing the Dice overlap between tissue masks, as formulated in Eq.~(\ref{eq:rigid_opt}):

\begin{equation}
\mathcal{T}^*
=
\arg\max_{\mathcal{T}}
\text{Dice}(M_S^\mathcal{T},M_T)
\label{eq:rigid_opt}
\end{equation}

To prevent unstable alignments, the estimated transformation is accepted only if the Dice overlap satisfies the threshold criterion defined in Eq.~(\ref{eq:dice_thresh}), otherwise, an identity transformation is retained.

\begin{equation}
\text{Dice}(M_S^{\mathcal{T}^*},M_T) \geq 0.7
\label{eq:dice_thresh}
\end{equation}

\subsubsection{Feature-based Refinement Using XFeat}

Although tissue morphology provides stable global initialization, mask-based alignment alone cannot recover fine structural correspondence. We therefore refine the rigid transformation using XFeat, a lightweight feature extraction framework optimized for precise correspondence estimation. XFeat extracts sparse and semi-dense keypoints from downsampled WSIs, enabling robust feature matching under substantial staining variability. Up to 16,000 keypoints are detected and matched between source and target images. Putative correspondences are filtered using geometric consistency constraints and validated through RANSAC optimization with approximately 1,000--2,000 iterations at 99.9\% confidence. To select the final rigid transformation, candidate alignments from TriMorph and XFeat are evaluated using normalized gradient fields (NGF), which measure structural similarity through image gradients. This formulation is particularly effective for multimodal histopathology registration. The NGF criteria is defined as:

\begin{equation}
\text{NGF}(I_T,I_S,\mathcal{T})
=
h^2
\sum_{i=1}^{N}
\left[
1 -
\left(
\frac{
\langle \nabla I_S(\mathcal{T}(x_i)), \nabla I_T(x_i)\rangle + \epsilon^2
}{
\|\nabla I_S(\mathcal{T}(x_i))\|_\epsilon
\|\nabla I_T(x_i)\|_\epsilon
}
\right)^2
\right]
\label{eq:ngf}
\end{equation}

where $x_i$ denotes image coordinates and $\epsilon$ controls robustness to weak gradients and noise.

% Given the candidate transformations produced by TriMorph and XFeat, the final rigid alignment is selected using the normalized gradient fields (NGF) metric, which evaluates structural consistency based on image gradients. The transformation minimizing NGF energy is selected as the final coarse rigid initialization.}

\subsubsection{Coarse Non-Rigid Registration}

Following rigid alignment, we perform coarse non-rigid registration to estimate large-scale tissue deformations between the source WSI $S$ and target WSI $T$. This stage compensates for nonlinear distortions, tissue stretching, compression, and other regional deformations that cannot be resolved through rigid transformations alone. To robustly model these variations while maintaining computational efficiency, we employ a hierarchical multi-resolution deformation framework inspired by both learning-based registration methods, including VoxelMorph~\cite{balakrishnan_voxelmorph_2018}, DeeperHistReg~\cite{wodzinski_regwsi_2024}, and classical deformable registration approaches such as ANTs~\cite{avants_reproducible_2011}.

We construct a multi-resolution image pyramid, starting from low-resolution representations to capture large-scale deformations and progressively refining the deformation field up to 1.25$\times$ resolution. At each pyramid level, spatial correspondence between the source and target images is modeled using a dense displacement field defined on a regular grid. The deformation field $\mathbf{D}$ is estimated by optimizing an energy function that balances appearance consistency and spatial regularity:

% 
% A multi-resolution image pyramid is employed, beginning from low-resolution representations to estimate large-scale deformations and progressively refining the displacement field till 1.25$\times$ resolution. At each pyramid level, spatial correspondence between the source and target is modeled using a dense displacement field defined over a regular grid.
% The displacement field is estimated by optimizing an objective function that balances \textbf{structural similarity} and \textbf{deformation smoothness}. Structural correspondence is measured using \textbf{local normalized cross-correlation (NCC)}, which provides robustness against local intensity variations across stains. Smoothness is enforced using a homogeneous regularization term that penalizes abrupt spatial variations in the displacement field:}

\begin{equation}
R(\mathbf{D}) = \lambda_s \left\|\nabla \mathbf{D} \right\|^2_{\text{Frob}}
\label{eq:deform}
\end{equation}

where:
\begin{itemize}
    \item $\mathbf{D}$ denotes the displacement field,
    \item $\nabla \mathbf{D}$ denotes the spatial gradient of the displacement field,
    \item $\left\|\cdot\right\|_{\text{Frob}}$ denotes the Frobenius norm, and
    \item $\lambda_s$ controls the strength of smoothness regularization.
\end{itemize}

Optimization is performed using the Adam optimizer, and the displacement field is progressively refined across six pyramid levels up to resolution $1.25\times$. This stage provides a more consistent coarse alignment and reduces the search space for the subsequent nuclei-based fine registration stage. For deformation estimation, local normalized cross correlation (NCC) with a window size of 7 is employed as the primary similarity metric during optimization:

\begin{equation}
\text{NCC} =
\frac{
\sum_{i=1}^{N} (T(x_i)-\bar{T})(S(x_i)-\bar{S})
}{
\sqrt{
\sum_{i=1}^{N}(T(x_i)-\bar{T})^2
\sum_{i=1}^{N}(S(x_i)-\bar{S})^2
}}
\label{eq:ngf}
\end{equation}

where $T$ and $S$ represent the target and source WSIs, respectively, $\bar{T}$ and $\bar{S}$ denote their respective mean intensities, and $N$ indicates the number of pixels within the overlap region. Table~\ref{tab:nonrigid} summarizes all parameters used for estimating the coarse-level deformation field.
\subsection{Fine-Grained  Registration (Cell-level Alignment)}

Coarse registration estimates large global tissue specific shifts, however residual misalignment persists at the cellular level, particularly in regions requiring accurate nuclei-level correspondence across WSIs. To address this, we introduce a fine-grained shape-aware point-set registration stage that operates on nuclei centroids extracted from both source and target WSIs. We formulate this stage as a shape-constrained registration problem that explicitly models the spatial organization of nuclei, enabling robust alignment under cross-stain and cross-section variability.

\begin{figure}[h!]
    \centering
    \includegraphics[width=1\textwidth]{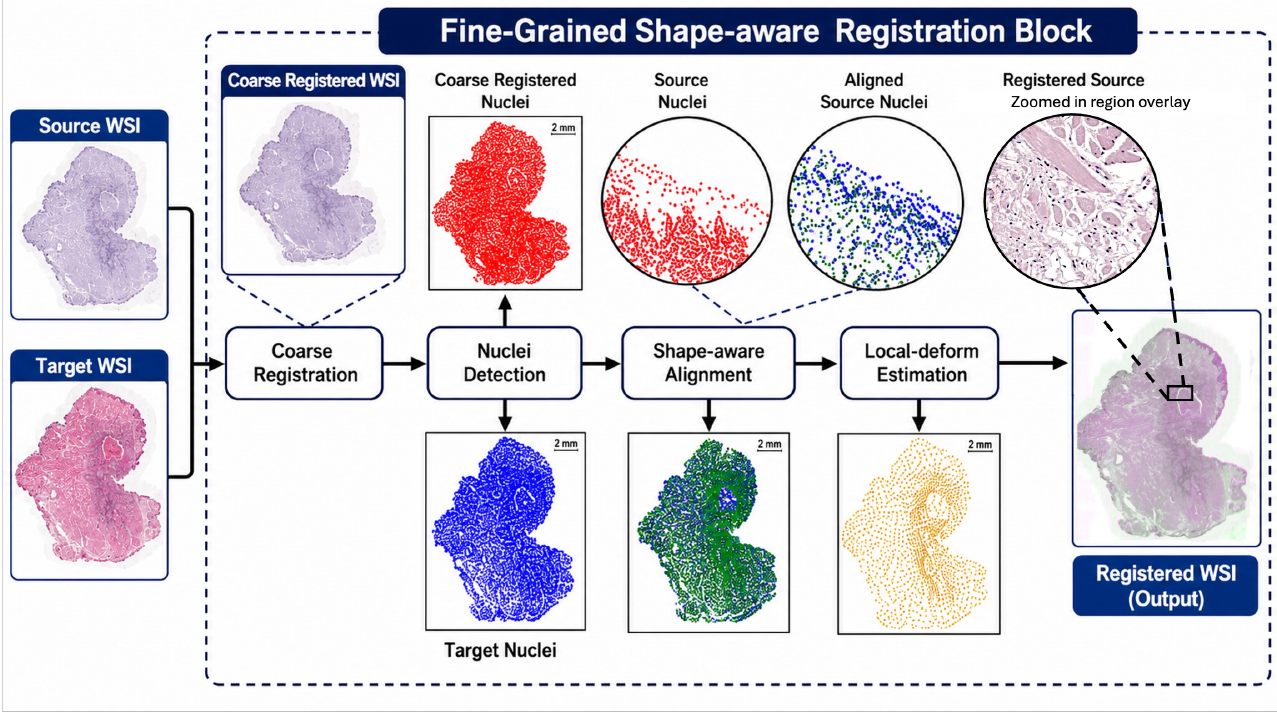}
    \captionsetup{width=\textwidth}
    \caption{Proposed fine shape-aware nuclei point-set registration. The coarse displacement field is first applied to the source WSI, producing a coarsely registered slide. Nuclei point sets are then detected from the target and coarsely registered WSIs, followed by shape-aware point-set alignment and local deformation estimation.}
    \label{fig:fine_shape}
\end{figure}

The fine registration stage operates on WSIs at $20\times$ or $40\times$ magnification and consists of three sequential components:
\begin{enumerate}
    \item nuclei point-set detection,
    \item shape-aware global point-set alignment, and
    \item local non-rigid refinement using CPD.
\end{enumerate}
A schematic overview of fine-grained pointset registration is shown in Figure~\ref{fig:fine_shape}.
\subsubsection{Nuclei Point-set Extraction}

To construct fine-scale correspondence, nuclei are represented as sparse point sets derived from centroid locations. Each WSI is partitioned into non-overlapping patches of size $1000 \times 1000$ pixels at full-resolution. Patches are converted to grayscale and processed using adaptive thresholding based on local intensity statistics. H-maxima and H-minima transforms are subsequently applied to identify candidate nuclear regions as local intensity extrema.
These markers are used in marker-controlled watershed segmentation applied to the distance transform of the binary nuclei mask, enabling robust separation of clustered nuclei while preserving local morphology. Finally, nuclei centroids are extracted and mapped to global WSI coordinates to ensure spatial consistency across the slide. The resulting source and target nuclei point sets are represented in Eq. (\ref{eq:source_point}), (\ref{eq:target_point}):

\begin{equation}
P_S = \{p_i\}_{i=1}^{n}
\label{eq:source_point}
\end{equation}

and

\begin{equation}
P_T = \{q_j\}_{j=1}^{m}
\label{eq:target_point}
\end{equation}

for the source and target WSIs, respectively.

\subsubsection{Shape-aware Point-set Registration}
Given the extracted nuclei point sets, fine registration is formulated as estimation of a rigid transformation that minimizes spatial and morphological discrepancy between source and target nuclei distributions. Source and target nuclei centroids are represented in Eq. (\ref{eq:source_p}), (\ref{eq:target_p}):

\begin{equation}
p_i = (x_i,y_i) \in \mathbb{R}^2
\label{eq:source_p}
\end{equation}

and

\begin{equation}
q_j = (x_j,y_j) \in \mathbb{R}^2
\label{eq:target_p}
\end{equation}

The transformation is parameterized by:

\begin{equation}
\theta = [w_x,w_y,\phi,s_c]
\label{eq:transform}
\end{equation}

Eq. (\ref{eq:transform}) represents translation ($w_x,w_y$), rotation $\phi$, and isotropic scaling $s_c$.

The transformation function is defined in Eq. \ref{eq:shape_transform}:

\begin{equation}
W_\theta(p)
=
s_c
\begin{bmatrix}
\cos\phi & -\sin\phi \\
\sin\phi & \cos\phi
\end{bmatrix}
p
+
\begin{bmatrix}
w_x \\
w_y
\end{bmatrix}
\label{eq:shape_transform}
\end{equation}

where $W_\theta(p)$ denotes the transformed position of source point set. To incorporate morphological consistency, each nucleus is additionally associated with a scalar shape descriptor derived from nuclear area. Shape descriptors are normalized across source and target point sets. We have defined shape descriptor as $\hat d_i$ in Eq. (\ref{eq:shape_norm}):

\begin{equation}
\hat{d}_i
=
\frac{
d_i - \min(d)
}{
\max(d)-\min(d)
}
\label{eq:shape_norm}
\end{equation}
We define a hybrid shape-aware distance metric:

\begin{equation}
d_\theta(p_i,q_j)
=
(1-\delta)
\|W_\theta(p_i)-q_j\|_2
+
\delta
|\hat{d}_i-\hat{d}_j|
\label{eq:shape_distance}
\end{equation}

where, in Eq. (\ref{eq:shape_distance}), $\delta \in [0,1]$ controls the relative contribution of spatial geometry and morphological consistency. The final optimization objective is defined in Eq. (\ref{eq:shape_energy}):

\begin{equation}
E(\theta)
=
\frac{1}{n}
\sum_{i=1}^{n}
\min_j
d_\theta(p_i,q_j)
\label{eq:shape_energy}
\end{equation}

For estimating nearest neighbor correspondence we have used k-d tree algorithm. Optimization is performed using Powell's derivative-free method due to the non-convex and non-differentiable nature of correspondence assignment. Optimization terminates when:

\begin{equation}
\|\theta^{i+1}-\theta^i\|_2 < \epsilon
\label{eq:terminate}
\end{equation}
or the maximum iteration limit is reached in Eq. (\ref{eq:terminate}).

To improve convergence stability, a progressive sampling strategy is adopted. For re-stained WSIs, full nuclei sets are used directly since tissue topology is largely preserved. For consecutive sections, optimization is initialized using 500 nuclei and progressively increased up to $2 \times 10^5$ nuclei points, with the optimal transformation propagated across sampling stages. Table~\ref{tab:algoparams} summarizes all parameters used for shape-aware point set registration.

\subsubsection{CPD-based Local Deformation Refinement}

Following rigid cellular registration, the final stage of CORE accounts for residual tissue distortions that cannot be captured by a global transformation. While shape-aware matching provides robust nuclei correspondences, neighboring cells may still exhibit spatial inconsistencies introduced during tissue sectioning and slide preparation. A non-linear deformation field is therefore estimated to further improve correspondence accuracy and preserve tissue structure. To correct these residual cell-scale misalignments, CORE performs a final local refinement using \emph{CPD}. It operates directly on the matched nuclei and estimates a smooth non-rigid transformation that aligns the source and target point sets while preserving the spatial relationships between neighboring cells. To improve robustness, the optimization is initialized using mutual nearest neighbor (MNN) correspondences, where a match is retained only if both nuclei are reciprocally identified as nearest neighbors. This bidirectional consistency criterion suppresses ambiguous matches and provides reliable anchors for deformation estimation. In practice, approximately $|C| \approx 5000$ MNN correspondences provide an effective balance between spatial coverage and computational efficiency. The resulting sparse displacement vectors are interpolated to generate a dense deformation field over the tissue domain and subsequently regularized using gaussian smoothing to suppress high-frequency artifacts. Transformation validity is further assessed using the jacobian determinant, with regions exhibiting non-positive values regularized to prevent local folding and ensure invertibility. The final deformation field is then applied to the source WSI, enabling accurate cell-level alignment while preserving the global tissue correspondence established in the preceding stages. The parameters used in the CPD refinement stage are summarized in Table~\ref{tab:cpd}.

\section{Datasets and Evaluation Metrics}

We evaluate CORE on \textbf{six} multimodal histopathology datasets spanning \textbf{30} distinct stain combinations across bright-field and immunofluorescence imaging modalities. The datasets include both consecutive and re-stained WSI pairs with substantial variations in tissue morphology, stain appearance, and acquisition protocols, enabling comprehensive evaluation under cross-stain registration settings.

\subsection{Datasets}

\begin{figure}[H]
    \centering
    \includegraphics[width=1\textwidth]{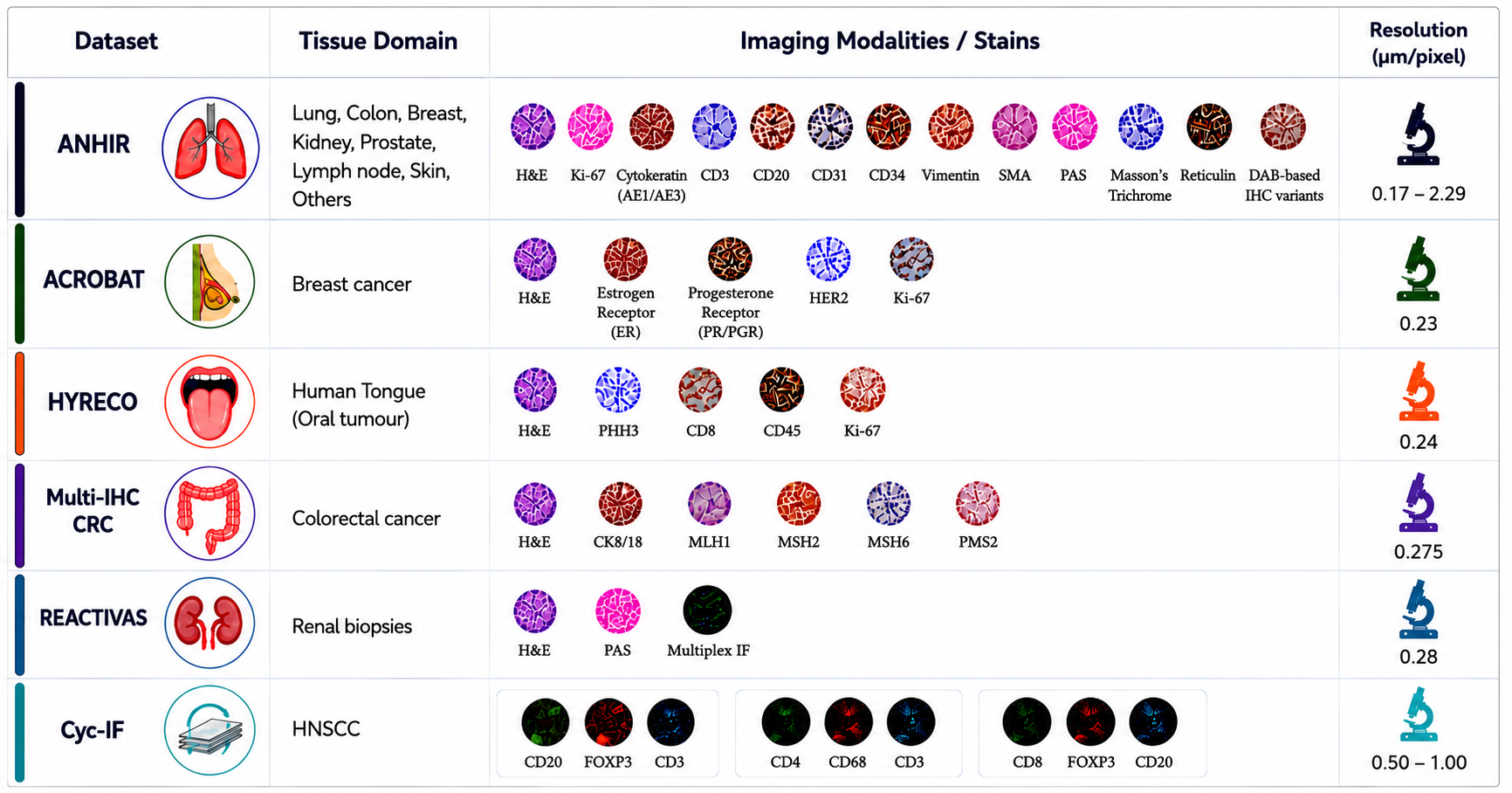}
    \captionsetup{width=\textwidth}
   \caption{ Datasets used for CORE for multi-stain WSI registration. The figure summarizes key characteristics of each dataset, including tissue types, staining modalities, and image resolution (µm/pixel).}
    \label{fig:dataset}
\end{figure}

We evaluate CORE on four public datasets and two private clinical datasets: ACROBAT \cite{weitz_acrobat_2024}, ANHIR \cite{borovec_anhir_2020}, HyReCo~\cite{van_der_laak_hyreco_2021}, Cyc-IF \cite{gatenbee_source_2023}, Multi-IHC CRC \cite{awan_deep_2023}, and REACTIVAS~\cite{web:reactivas}. Together, these datasets cover H\&E, IHC, PAS, Cyc-IF, and mIF imaging modalities and present varying levels of structural deformation and modality-specific appearance variations, providing a comprehensive benchmark for registration evaluation. An overview of dataset characteristics is provided in Figure~\ref{fig:dataset}.

\subsubsection{ACROBAT-AutomatiC Registration Of Breast cAncer Tissue }
% The ACROBAT dataset \cite{weitz_acrobat_2024} consists of consecutive breast cancer WSIs stained with H\&E, ER, Ki67, PGR, and HER2. It includes 750 training, 100 validation, and 303 test cases, totaling 3,406 WSIs. The dataset provides clinically relevant cross-marker variability within a consistent tissue domain.

The ACROBAT dataset \cite{weitz_acrobat_2024} is a large-scale benchmark for WSI registration in breast cancer histopathology. It consists of consecutive tissue sections from breast cancer patients stained with H\&E and multiple immunohistochemical markers, including ER, Ki67, PGR, and HER2.
The dataset contains 750 training, 100 validation, and 303 test cases, corresponding to 4,212 WSIs from routine clinical workflows. Images are collected across multiple scanners and staining conditions, enabling evaluation under diverse setting.
ACROBAT was introduced to benchmark both rigid and non-rigid WSI registration methods using H\&E-to-IHC, with landmark correspondence provided through expert-annotated landmarks.
\subsubsection{ANHIR - Automatic Non-rigid Histological Image Registration } 

The ANHIR dataset \cite{borovec_anhir_2020} is a benchmark for non-rigid registration of histological images in multiple types of tissues and staining protocols. It contains 481 image pairs (230 training and 251 evaluation) that spans 8 tissue types and 18 different stains. The dataset was designed to evaluate deformable registration methods under challenging conditions, including large inter-slide deformations, heterogeneous staining appearance, and absence of strict structural correspondence. ANHIR is widely used as a standard benchmark for evaluating the robustness and generalization of histology registration algorithms.

\subsubsection{HyReCo - Hybrid Re-stained and Consecutive Histological Serial Sections}

The HyReCo dataset comprises two complementary subsets. \textit{Subset A} (consecutive section slides) contains 9 groups of consecutive tissue sections stained with H\&E, CD8, CD45, and Ki67. PHH3 stained slides were generated by bleaching and re-staining the original H\&E sections using a t-Cyc-IF (tissue-based Cyclic Immunofluorescence) like technique. Each section includes 11--19 manually verified landmarks, resulting in 138 landmarks per stain and 690 in total. \textit{Subset B} (re-stained slides) consists of 54 paired H\&E--PHH3 images, containing a total of 2,303 annotated landmarks (approximately 43 per pair).

\subsubsection{Cyc-IF -- Cyclic Multiplex Immunofluorescence}

The Cyc-IF dataset consists of 9 mIF WSIs acquired across multiple staining cycles. Each cycle captures different biomarker panels, including (CD20, FOXP3, CD3), (CD4, CD68, CD3), and (CD8, FOXP3, CD20). 
This dataset is used to evaluate cross-cycle registration under challenging conditions, where consecutive imaging rounds introduce intensity variation, partial signal overlap, and potential spatial misalignment due to repeated staining and imaging procedures. These characteristics make Cyc-IF a challenging benchmark for robust mIF registration methods.

\subsubsection{Multi-IHC CRC}
The Multi-IHC CRC dataset, obtained from the University Hospitals Coventry and Warwickshire (UHCW) NHS Trust, comprises WSIs from 86 distinct patients \cite{awan_deep_2023}. Each case features 16 high-resolution slides scanned at 0.275 \textmu m/px  with various stains including CK8/18, mismatch repair (MMR) markers (MLH1, MSH2, MSH6, PMS2), and H\&E. For evaluation of registration method, we selected 7 cases with 6 slides per case for direct comparison with Awan et al.~\cite{awan_deep_2023}. 

\subsubsection{REACTIVAS}

The REACTIVAS dataset \cite{web:reactivas} consists of renal biopsy WSIs stained with H\&E, PAS, and mIF protocols. We use 11 paired samples across all staining modalities for evaluation. The inclusion of mIF provides multi-channel molecular and cellular information that complements structural histology and introduces additional challenges for cross-modality registration due to increased appearance and representation variability.

\subsection{Implementation Details}

All algorithms were implemented in Python and executed on NVIDIA Tesla A100 GPUs with 32GB memory. WSI visualization and qualitative inspection of registration outputs are performed using TiaViz \cite{eastwood_tiaviz_2024}, integrated within the TIAToolbox \cite{pocock_tiatoolbox_2022} framework.

\subsection{Evaluation Metrics}

To quantitatively evaluate the performance of the proposed registration framework, we employed the Target Registration Error (TRE) and the relative Target Registration Error (rTRE), which are widely used metrics for landmark-based histopathology image registration evaluation.

\subsubsection{Target Registration Error (TRE)}

For each image pair $j$, the registration accuracy was measured using the Euclidean distance between corresponding landmarks in the target image and the warped source image, as defined in Eq.~(\ref{eq:tre}):

\begin{equation}
\mathrm{TRE}(T_j, W_j) = \left\| x_{l}^{T_j} - x_{l}^{W_j} \right\|_2
\label{eq:tre}
\end{equation}

where $T_j$ and $W_j$ denote the target and warped source images for image pair $j$, respectively. 
$x_{l}^{T_j}$ represents the landmark coordinate in the target image, while $x_{l}^{W_j}$ denotes the corresponding landmark coordinate in the warped source image.

\subsubsection{Relative Target Registration Error (rTRE)}

To make the registration error independent of image resolution and image size, the TRE values were normalized by the diagonal length of the target image. The relative TRE (rTRE) was computed as defined in Eq.~(\ref{eq:rtre}):

\begin{equation}
\mathrm{rTRE}(T_j, W_j) =
\frac{\mathrm{TRE}(T_j, W_j)}
{\sqrt{w_j^2 + h_j^2}}
\label{eq:rtre}
\end{equation}

where $w_j$ and $h_j$ denote the width and height of the target image, respectively, and $\sqrt{w_j^2 + h_j^2}$ corresponds to the image diagonal length. The formulation in Eq.~(\ref{eq:rtre}) generates a set of rTRE values for each image pair based on all annotated landmark correspondences.

\subsubsection{Dataset-Level Aggregation Metrics}

To provide robust evaluation statistics across the dataset, we aggregated the landmark-wise rTRE values using both median and maximum operators for each image pair. Dataset-level performance was then computed by averaging these aggregated values across all $N$ image pairs. The average median rTRE (Med-rTRE) was computed as defined in Eq.~(\ref{eq:med_rtre}):

\begin{equation}
\mathrm{Med\text{-}rTRE}
=
\frac{1}{N}
\sum_{j=1}^{N}
\mathrm{median}
\left(
\mathrm{rTRE}_j
\right)
\label{eq:med_rtre}
\end{equation}

Similarly, the average maximum rTRE (Max-rTRE) was computed as defined in Eq.~(\ref{eq:max_rtre}):

\begin{equation}
\mathrm{Max\text{-}rTRE}
=
\frac{1}{N}
\sum_{j=1}^{N}
\max
\left(
\mathrm{rTRE}_j
\right)
\label{eq:max_rtre}
\end{equation}

where $\mathrm{rTRE}_j$ represents the set of relative registration errors for all landmarks in image pair $j$, and $N$ denotes the total number of evaluated image pairs.

The median rTRE provides a robust estimate of overall registration accuracy, while the maximum rTRE reflects the worst-case local misalignment for each image pair.
Together, these datasets and evaluation metrics enable comprehensive assessment of registration robustness across diverse tissue types, staining protocols, and multimodal WSI acquisition settings. Table~\ref{tab:abb} presents the abbreviations and definitions of all metrics used in the evaluation. The choice of metric used in our analysis is based on the metric used in the original papers \cite{wodzinski_regwsi_2024, lotz_comparison_2023, awan_deep_2023, borovec_anhir_2020} of the respective datasets for a fair comparison.
\vspace{0.5em}

\section{Results}
We evaluate CORE against state-of-the-art registration methods across six multimodal datasets and perform ablation analysis to assess each design component. As most existing WSI registration approaches operate at coarse resolution, we report results at both coarse and fine stages. The coarse stage enables fair comparison with prior work, while the fine stage demonstrates the additional precision provided by cell-centric refinement. Fine registration is not reported on ACROBAT and Cyc-IF dataset at higher magnification due to its limited resolution 10$\times$ of images in the dataset.

\subsection{ACROBAT}
\begin{figure}[H]  % Also force top placement
    \centering
    \includegraphics[width=1.02\textwidth]{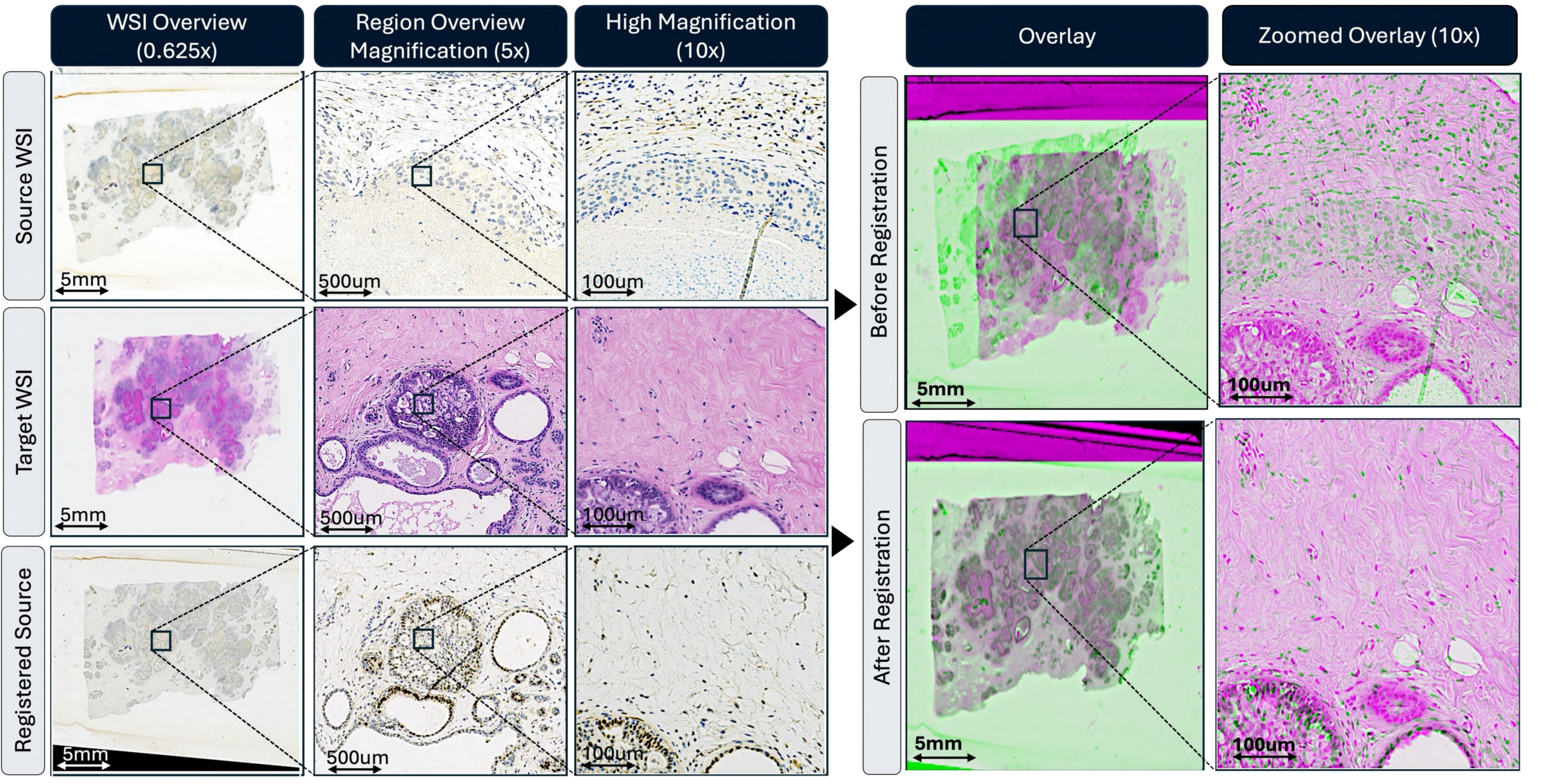}
     \captionsetup{width=\textwidth}
     \caption {CORE results on the ACROBAT dataset. WSI overview (0.625$\times$): shows target (H\&E), source (ER), and registered source WSIs. Region overview (5$\times$): corresponding tissue regions across the images. High magnification (10$\times$): zoomed-in local regions showing fine structural details. Registration overlays: alignment of target with source and target with registered source at different scales.}
    \label{fig:result_acrobat}
\end{figure}

ACROBAT dataset is the most recent and largest publicly available WSI registration dataset released as part of the grand challenge. Table~\ref{tab:acrobat} compares CORE against ten state-of-the-art coarse registration methods on the ACROBAT dataset~\cite{weitz_acrobat_2024}. We report the Average Median Target Registration Error (AMTRE), the primary metric defined by the challenge organizers, to ensure a direct and fair comparison with existing methods under a consistent evaluation protocol. CORE achieves the lowest median AMTRE at the 90th percentile (139.0~$\mu$m), reducing registration error compared to state-of-the-art methods, as well as significantly reducing runtime (14s vs.\ 50--120s for most competing methods). This demonstrates a strong accuracy–efficiency trade-off compared to existing approaches. 
Qualitative results in Figure~\ref{fig:result_acrobat} show that CORE reduces global misalignment at low magnification and preserves structural consistency after registration. Additionally, Figure~\ref{fig:acrobat_reg_artifacts} shows the performance of CORE registration method in the presence of artifacts.

\begin{table}[H]
\footnotesize
\centering
\captionsetup{width=\textwidth}
\caption{Performance comparison of coarse registration methods on the \textbf{ACROBAT} dataset. Results are reported as median AMTRE (90th percentile, $\mu$m) $\pm$ standard deviation, along with mean runtime (s). Lower AMTRE values indicate better performance.}

\renewcommand{\arraystretch}{1.2}
\setlength{\tabcolsep}{5pt}

\begin{tabular}{p{1.9cm} p{3cm} p{4.2cm} p{1.8cm}}

\toprule
\textbf{Resolution} &
\textbf{Method} &
\textbf{AMTRE-Percentile90 [$\mu$m] $\pm$ std} &
\textbf{Mean Time (s)} \\
\midrule

\multirow{11}{*}{\textbf{Coarse}}

& Initial
& 8556.00 $\pm$ 17601.43
& - \\

& DFBR \cite{awan_deep_2023}
& 1447.36 $\pm$ 13964.30
& 50 \\

& DFReg \cite{kim2025distortion}
& 276.80 $\pm$ 1287.90
& 52 \\

& MEDAL
& 248.54 ± 4771.10
& - \\

& NEMESIS \cite{wolterink2022implicit}
& 209.76 ± 6554.51
& 120 \\

& VALIS \cite{gatenbee_virtual_2023}
& 176.70 $\pm$ 6986.91
& 39 \\

& Gestalt \cite{gestalt_diagnostics}
& 149.08 $\pm$ 4144.50
& - \\

& AGH \cite{wodzinski_deephistreg_2021}
& 141.77 $\pm$ 6714.37
& - \\

& HistokatFusion \cite{lotz_comparison_2023}
& 141.64 $\pm$ 2953.77
& 50 \\

& DeeperHistReg \cite{wodzinski_regwsi_2024}
& 140.33 $\pm$ 6707.98
& 70 \\

& \textbf{CORE}
& \textbf{139.00$\pm$7196.80}
& \textbf{14} \\

\bottomrule
\end{tabular}

\label{tab:acrobat}
\end{table}

\subsection{ANHIR}

ANHIR dataset serves as a benchmark for evaluating the non linear registration performance. Since the dataset is most diverse including different tissues and staining variations. It also has challenging cases including severe tissue deformations missing tissues and staining artifacts. Table~\ref{tab:anhir} compares CORE against eleven state-of-the-art methods on the ANHIR dataset~\cite{borovec_anhir_2020}. At the coarse stage, CORE achieves the best overall performance in terms of rTRE, obtaining the lowest AArTRE (0.0040), MArTRE (0.0026), and MMrTRE (0.0015). The subsequent fine stage further reduces error metrics, reducing AArTRE to 0.0034 and MMrTRE to 0.0012, thereby demonstrating the effectiveness of the proposed cell-centric refinement strategy. Although the fine stage increases the overall runtime (1680~s), it enables cellular-scale alignment that is critical for downstream applications such as spatial transcriptomics and single-cell analysis. We also provide registration results at each stage in Table~\ref{tab:anhir_step} for qualitative evaluation of improved performance of CORE from coarse to fine stage. Similarly, qualitative results in Figure~\ref{fig:result_anhir} further corroborate these findings, showing consistent alignment across resolutions, with fine-stage refinement substantially improving local structural correspondence at 40$\times$ magnification.

\begin{figure}[H]  % Also force top placement
    \centering
    \includegraphics[width=1.02\textwidth]{figure_7.pdf}
      \captionsetup{width=\textwidth}
  \caption{CORE results on the ANHIR dataset. WSI overview (0.625$\times$): shows low-magnification views of target (H\&E), source (ER), and registered WSIs. Region overview (5$\times$): intermediate-resolution tissue regions across target, source, and registered images. High magnification (40$\times$): full-resolution local regions highlighting fine structural details. Registration overlays: alignment of target-source and target-registered pairs before and after registration. Zoomed overlays (40$\times$): high-resolution visualization of registration quality at cellular scale.}
    \label{fig:result_anhir}
\end{figure}

\begin{sidewaystable}
\centering
\captionsetup{width=\textheight}
\caption{Performance comparison of \textbf{CORE} coarse and fine methods with existing registration methods on the \textbf{ANHIR} dataset using relative Target Registration Error (rTRE). Results are presented with two aggregation levels: \textbf{Average} and \textbf{Median} across all image pairs. Methods are sorted by Average rTRE (AArTRE). The proposed \textit{fine} method achieves the lowest error in both AArTRE and MMrTRE metrics. Best value per column is shown in \textbf{bold}.} \footnotetext{Standard deviation is computed from the same underlying error distribution as the reported rTRE metrics and is shown once per method for average, median and max to avoid redundancy.}
\setlength{\tabcolsep}{3pt}
\renewcommand{\arraystretch}{1.2}

\begin{tabular}{@{}p{2.1cm}p{3.2cm}p{3cm}p{2cm}p{3.0cm}p{2.0cm}p{3.0cm}p{2.0cm}p{2cm}@{}}
\toprule
\textbf{Resolution} & \textbf{Method} & \multicolumn{2}{c}{\textbf{Average rTRE}} & \multicolumn{2}{c}{\textbf{Median rTRE}} & \multicolumn{2}{c}{\textbf{Max rTRE}} & \textbf{Mean Time (s)} \\
\cmidrule(lr){3-4} \cmidrule(lr){5-6} \cmidrule(lr){7-8}
& & \textbf{AArTRE} & \textbf{AMrTRE} & \textbf{MArTRE} & \textbf{MMrTRE} & \textbf{AMxrTRE} & \textbf{MMxrTRE} & \\
\midrule
\multirow{13}{*}{\textbf{Coarse}}
& Initial                                       & 0.1340 $\pm$ 0.1279 & 0.0684  & 0.1354 $\pm$ 0.1311 & 0.0665  & 0.2338 $\pm$ 0.2030 & 0.1157  & -- \\
& NiftyReg \cite{rueckert2002nonrigid}                   & 0.1120 $\pm$ 0.0898 & 0.0372  & 0.1136 $\pm$ 0.0903 & 0.0355  & 0.2010 $\pm$ 0.0640 & 0.0714  & 0.14 \\
& bUnwarpJ \cite{arganda2006consistent}                   & 0.1097 $\pm$ 0.0907 & 0.0290  & 0.1105 $\pm$ 0.0908 & 0.0260  & 0.1995 $\pm$ 0.0737 & 0.0727  & 10.57 \\
& Elastix \cite{klein2009elastix}                     & 0.0964 $\pm$ 0.0310 & 0.0074  & 0.0956 $\pm$ 0.0310 & 0.0054  & 0.1857 $\pm$ 0.0283 & 0.0353  & 210 \\
& ANTs~\cite{avants_reproducible_2011} & 0.0991 $\pm$ 0.0142 & 0.0072  & 0.0992 $\pm$ 0.0143 & 0.0058 & 0.1861 $\pm$ 0.0206 & 0.0531  & 2894.5 \\
& DROP \cite{glocker2008dense}                       & 0.0861 $\pm$ 0.1415 & 0.0042 & 0.0867 $\pm$ 0.1450 & 0.0028 & 0.1644 $\pm$ 0.1796 & 0.0273 & 239.4 \\
& DFBR~\cite{awan_deep_2023}                    & 0.0055 $\pm$ 0.0079 & 0.0029 & 0.0040 $\pm$ 0.0075 & 0.0018  & 0.0275 $\pm$ 0.0240 & 0.0203  & 240 \\
& AGH~\cite{wodzinski_multistep_2020}           & 0.0053 $\pm$ 0.0294 & 0.0032  & 0.0036 $\pm$ 0.0305 & 0.0019  & 0.0283 $\pm$ 0.0418 & 0.0225  & 393 \\
& DeeperHistReg~\cite{wodzinski_regwsi_2024}    & 0.0044 $\pm$ 0.0216 & 0.0029  & 0.0029 $\pm$ 0.0211 & 0.0017  & 0.0280 $\pm$ 0.0201 & 0.0225  & 120.00 \\
& HistokatFusion~\cite{lotz_comparison_2023}    & 0.0044 $\pm$ 0.0195 & \textbf{0.0027}  & 0.0029 $\pm$ 0.0206 & 0.0018  & 0.0251 $\pm$ 0.0222 & 0.0188  & \textbf{9.6} \\
& CKVST                                         & 0.0043 $\pm$ 0.0489 & 0.0032  & 0.0027 $\pm$ 0.0506 & 0.0023  & \textbf{0.0239} $\pm$ 0.0209 & 0.0189  & 468 \\
& UPENN \cite{venet2019accurate}                                        & 0.0042 $\pm$ 0.0034 & 0.0029  & 0.0029 $\pm$ 0.0026 & 0.0019  & \textbf{0.0239} $\pm$ 0.0198 & 0.0190  & 960 \\
& \textbf{CORE}                          & \textbf{0.0040} $\pm$ 0.0361 & 0.00278  & \textbf{0.0026} $\pm$ 0.0370 & \textbf{0.0015}  & 0.0240 $\pm$ 0.0577 & \textbf{0.0187} & 12 \\
\midrule
\textbf{Fine} & \textbf{CORE}            & \textbf{0.0034} $\pm$ 0.0122 & \textbf{0.0021}  & \textbf{0.00245} $\pm$ 0.0092 & \textbf{0.0012}  & \textbf{0.0223} $\pm$ 0.0176 & \textbf{0.0180}  & 1680 \\
\bottomrule
\end{tabular}
\label{tab:anhir}
\end{sidewaystable}

\subsection{HyReCo}
\begin{figure}[H]  % Also force top placement
    \centering
    \includegraphics[width=1.02
    \textwidth]{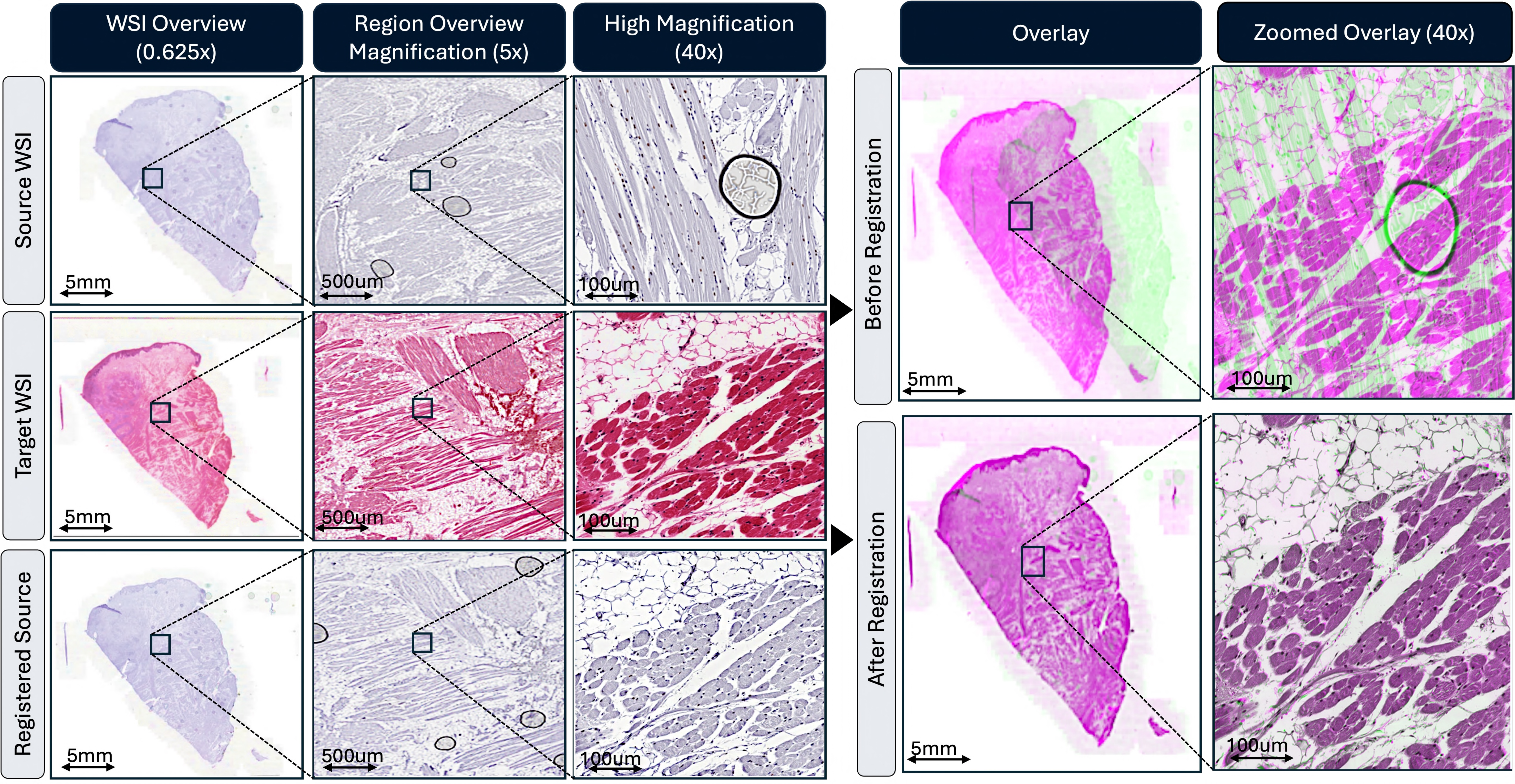}
    \captionsetup{width=\textwidth}
       \caption{CORE results on the HyReCo dataset.  WSI overview (0.625$\times$): low-magnification views of target (H\&E), source (PHH3), and registered WSIs. Region overview (5$\times$): intermediate-resolution tissue regions across target, source, and registered images. High magnification (40$\times$): full-resolution local regions highlighting fine structural details. Registration overlays: comparison of target-source before registration, and target-registered alignments. Zoomed overlays (40$\times$): high-resolution visualization of registration quality at cellular scale.}
    \label{fig:result_hyreco}
\end{figure}

HyReCo dataset serves as a benchmark for evaluating CORE performance on both consecutive and re-stained sections. Table~\ref{tab:hyreco-nonrigid} shows quantitative comparison of CORE with existing state-of-the-art models on both consecutive and re-stained sections. Similarly, Table~\ref{tab:hyreco-rigid} reports comparative analysis of existing feature extractors performance with CORE on coarse-stage registration. We have also reported TRE at each stage for demonstrating performance improvement from coarse to fine stages in Table~\ref{tab:hyreco_step}. CORE achieves state-of-the-art performance on both settings, reducing median TRE to 4.35~$\mu$m (consecutive) and 0.41~$\mu$m (re-stained). These results demonstrate robustness under both mild and strong staining transformations. The fine stage further improves alignment accuracy, particularly on re-stained pairs where structural correspondence is highly preserved. Despite higher computational cost, the fine stage provides consistent gains by refining residual local deformations not addressed by coarse registration. Qualitative results in Figure~\ref{fig:result_hyreco} shows visual registration performance of CORE across coarse to fine resolutions and staining conditions.

\begin{table}[H]
\footnotesize
\centering
\captionsetup{width=\textwidth}
\caption{Performance comparison of \textbf{CORE}  method on the \textbf{HyReCo} Dataset (consecutive and re-stained Histological Sections) in $\mu$m. The proposed \textit{fine} method achieves high precision in comparison with state-of-the-art for both consecutive and re-stained tissue types.}
\renewcommand{\arraystretch}{1.2}
\begin{tabular}{
p{1.5cm}
p{3.3cm}
p{2.2cm}
p{1.8cm}
p{2.2cm}
p{1.8cm}
}
\toprule
\multirow{2}{*}{\textbf{Resolution}} &
\multirow{2}{*}{\textbf{Technique}} &
\multicolumn{2}{c}{\textbf{Consecutive Sections}} &
\multicolumn{2}{c}{\textbf{Re-stained Sections}} \\
& &
\textbf{AMTRE [$\mu$m]} &
\textbf{Average Time (s)} &
\textbf{AMTRE [$\mu$m]} &
\textbf{Average Time (s)} \\
\midrule
\multirow{6}{*}{\textbf{Fine}}
& Initial & 2583.96 & -- & 118.95 & -- \\
% & Superpoint+LightGlue \cite{kim2025distortion} }& 43.76} & 0.114}  & -  & - \\
& DFReg \cite{kim2025distortion} & 35.10 & 0.327 &  -& - \\
& HistokatFusion \cite{lotz_comparison_2023} & 5.30 & 1831 & 0.90 & 1831 \\
& DeeperHistReg \cite{wodzinski_regwsi_2024} & 4.96 & 600 & 0.59 & 600 \\
& Elastix \cite{klein2009elastix} & 17.89 &
387 &
5.32 &
387 \\
& VALIS \cite{gatenbee_virtual_2023} &
15.52 &
460 &
2.50 &
460 \\
& \textbf{CORE} &
\textbf{4.35 $\pm$ 3.671} &
2850 &
\textbf{0.41 $\pm$ 1.272} &
2850 \\
\bottomrule
\end{tabular}
\label{tab:hyreco-nonrigid}
\end{table}
\footnotetext{In Table 3, standard deviations for several competing methods are not reported in the original publications. For consistency, we report standard deviation only for the proposed CORE method where applicable.}

\subsection{Cyc-IF}
\begin{figure}[H]  % Also force top placement
    \centering
    \includegraphics[width=1\textwidth]{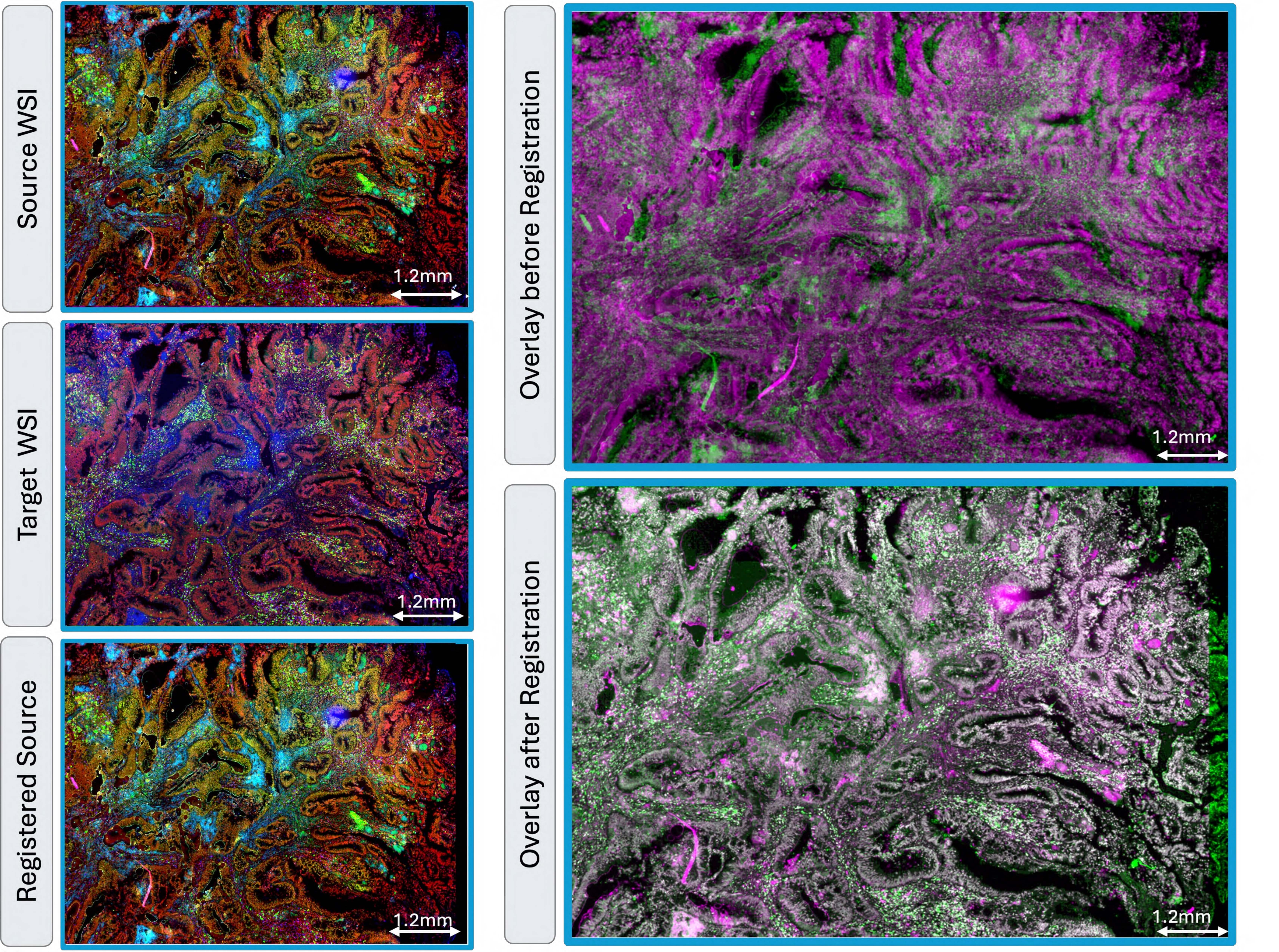}
    \captionsetup{width=\textwidth}
\caption{ Visualization of Cyc-IF images before and after CORE registration. Registration is performed between source pair $\rightarrow$ (CD20 FOXP3 CD3), and target pair$\rightarrow$  (CD4 CD68 CD3) .} 
    \label{fig:result_cycif}
\end{figure}

To evaluate the effectiveness and efficiency of CORE in an immunofluorescence-only setting, we perform experiments on the Cyc-IF dataset. Cyc-IF consists of mIF samples acquired across multiple staining cycles, including marker panels (CD20, FOXP3, CD3), (CD4, CD68, CD3), and (CD8, FOXP3, CD20). This setting is particularly challenging due to strong inter-cycle appearance variations, limited structural consistency, and potential spatial distortions introduced during repeated staining and imaging. These factors make reliable cross-cycle alignment difficult, and provide a rigorous testbed for evaluating robust multimodal registration methods. Table \ref{tab:cycif_registration} shows result of CORE performance in comparison with state-of-the-art methods and despite modality variations our model outperforms the baseline. Figure \ref{fig:result_cycif} shows qualitative result of CORE registration on Cyc-IF dataset before and after registration.
\begin{table}[H]
\footnotesize
\centering

\captionsetup{width=\textwidth}
\caption{Performance comparison on the \textbf{Cyc-IF} dataset. Values are TRE in $\mu$m, aggregated over Cyc-IF panels. }
\label{tab:cycif_registration}
\renewcommand{\arraystretch}{1.2}
\setlength{\tabcolsep}{6pt}
\begin{tabular}{p{1.8cm}  p{2.9cm} p{2.3cm} p{2.3cm} p{1.8cm}}
\toprule
\textbf{Resolution} &
\textbf{Method} &
\textbf{AMTRE [$\mu$m]} &
\textbf{AMxTRE [$\mu$m]} &
\textbf{Avg Time (s)} \\
\midrule
\multirow{3}{*}{Coarse}
& Initial & 134.11$\pm$41.66 & 175.77 & 0.01 \\
& SIFT \cite{lindeberg_scale_2012} & 52.4$\pm$0.15 &67.9 & 3.5\\
& VALIS \cite{gatenbee_virtual_2023} & 2.85$\pm$0.20 & 3.05 & 15.13 \\
& DeeperHistReg \cite{wodzinski_regwsi_2024} & 2.31$\pm$0.0012 &7.26 & 51.18  \\
& CORE & 1.76$\pm$0.0001 & 	2.7821& 4.76 \\
\bottomrule
\end{tabular}
\end{table}
\subsection{MULTI-IHC CRC}

Following evaluation on public datasets, CORE  was further tested on private datasets for evaluating model generalizability across diverse staining protocols compared to stains included in the public datasets. Table~\ref{tab:comet} evaluates CORE on the Multi-IHC CRC dataset, representing diverse immunohistochemical staining protocols. CORE achieves the lowest registration error (AArTRE: 0.0043) while maintaining efficient runtime (14.37 s), demonstrating strong generalization. The fine stage further improves performance to 0.00228 AArTRE, indicating robust refinement under heterogeneous IHC marker distributions. Overall, CORE consistently improves alignment accuracy while maintaining computational efficiency across clinically diverse cases. Table~\ref{tab:comet_step} summarizes step-wise results on the Multi-IHC CRC dataset, detailing performance before initial alignment, after coarse alignment, and following fine alignment. Figure~\ref{fig:result_comet} illustrates representative target (MSH2), source (MLH1) and registered slides from the Multi-IHC CRC dataset, along with the registration results overlay before and after alignment, at coarse and fine resolution.
\begin{table}[H]
\footnotesize
\centering
\captionsetup{width=\textwidth}
\caption{Performance comparison of \textbf{CORE} coarse and fine method on the \textbf{Multi-IHC CRC} Dataset in $\mu$m. Registration errors are reported in rTRE.}
\renewcommand{\arraystretch}{1.2}
\setlength{\tabcolsep}{5pt}
\begin{tabular}{
p{1.7cm}
p{2.5cm}
p{3.3cm}
p{2.0cm}
p{2.0cm}
p{1.8cm}
}
\toprule
\textbf{Resolution} &
\textbf{Method} &
\textbf{AArTRE [$\mu$m]} &
\textbf{AMrTRE [$\mu$m]} &
\textbf{AMxrTRE [$\mu$m]} &
\textbf{Avg Time (s)} \\
\midrule
\multirow{8}{*}{\textbf{Coarse}}
& Initial
& 0.1807
& 0.1520
& 0.3615
& -- \\
& VALIS
& 0.1359 $\pm$ 0.0954
& 0.1420
& 0.5637
& 85 \\
& PALOM \cite{lin_high-plex_2023}
& 0.1744 $\pm$ 0.0533
& 0.1748
& 0.1800
& 9.16  \\
% & 9.16 $\pm$ 0.98} \\
& ANTs \cite{avants_reproducible_2011}
& 0.0819 $\pm$ 0.1032
& 0.0623
& 0.1298
& 113.36 \\
% & 113.36 $\pm$ 24.18} \\
& DFBR \cite{awan_deep_2023}
& 0.0050 $\pm$ 0.0015
& 0.0031
& 0.0188
& 50.13 \\
& Elastix \cite{klein2009elastix}
& 0.0641 $\pm$ 0.0418
& 0.0642
& 0.0685
& 7.62  \\
% & 7.62 $\pm$ 1.82} \\
& DeeperHistReg \cite{wodzinski_regwsi_2024}
& 0.0049 $\pm$ 0.0018
& 0.0029
& 0.0196
& 70 \\
& \textbf{CORE}
& \textbf{0.0043 } {$\pm$ 0.0015}
& \textbf{0.0024}
& \textbf{0.0153}
& \textbf{14.37} \\
\midrule
\textbf{Fine}
& \textbf{CORE}
& \textbf{0.0022}{ $\pm$ 0.0013}
& \textbf{0.0011}
& \textbf{0.0142}
& \textbf{2172} \\
\bottomrule
\end{tabular}
\label{tab:comet}
\end{table}
\footnotetext{Standard deviation is computed from the same underlying error distribution as the reported TRE metrics and is shown once per method to avoid redundancy.}

\begin{figure}[H]  % Also force top placement
    \centering
    \includegraphics[width=1\textwidth]{figure_10.pdf}
          \captionsetup{width=\textwidth}
   \caption{CORE results on the Multi-IHC CRC dataset. WSI overview (0.625$\times$): low-magnification views of target (MLH1), source (MSH2), and registered WSIs. Region overview (5$\times$): intermediate-resolution tissue regions across target, source, and registered images. High magnification (40$\times$): full-resolution local regions showing fine structural details. Registration overlays: comparison of target before registration, target-source, and target-registered alignments. Zoomed overlays (40$\times$): high-resolution visualization of registration quality at cellular scale.}
    \label{fig:result_comet}
\end{figure}
\subsection{REACTIVAS}
REACTIVAS represents the most challenging cross-modality setting, including H\&E, PAS, and mIF imaging. We perform two systematic comparisons: H\&E-PAS and H\&E-mIF. Table~\ref{tab:reactivas} reports quantitative results across both modality pairs.  At the coarse stage, CORE achieves the lowest error on both PAS (1.03~$\mu$m) and mIF (2.05~$\mu$m), outperforming classical and deep learning baselines under strong modality shifts. The fine stage further reduces error to 0.36~$\mu$m and 0.85~$\mu$m, respectively, demonstrating strong robustness under extreme appearance differences. These improvements highlight the effectiveness of combining deep feature-based coarse alignment with cell-centric refinement for cross-modal registration. Table~\ref{tab:reactivas_step} reports a step-wise breakdown of CORE's pipeline stages (initial, coarse, and fine). Qualitative results in Figure~\ref{fig:result_reactivas} and Figure~\ref{fig:result_mIF} confirm consistent alignment across modalities, with fine-stage refinement improving local structural correspondence despite severe intensity and modality variations.  
Additionally, Figure~\ref{fig:reg_mIF} shows registration visualization on PAS--mIF WSIs in the presence of missing tissue artifacts showing robustness of our proposed coarse and fine models. 
\begin{table}[H]
\footnotesize
\centering
    \captionsetup{width=\textwidth}
\caption{Performance comparison of \textbf{CORE} coarse and fine method on the \textbf{REACTIVAS} (H\&E, PAS and mIF) dataset. Registration errors are reported in TRE.}
\renewcommand{\arraystretch}{1.2}
\begin{tabular}{p{2cm} p{3cm} p{2.4cm} p{1.7cm} p{2.4cm} p{1.7cm}}
\toprule
\multirow{2}{*}{\textbf{Resolution}} & \multirow{2}{*}{\textbf{Technique}} & \multicolumn{2}{c}{\textbf{PAS Sections}} & \multicolumn{2}{c}{\textbf{mIF Sections}} \\
\cmidrule(lr){3-4} \cmidrule(lr){5-6}
& & \textbf{AMTRE [$\mu$m]} & \textbf{Avg Time(s)} & \textbf{AMTRE [$\mu$m]} & \textbf{Avg Time(s)} \\
\midrule
\multirow{3}{*}{\textbf{Coarse}} 
& Initial  & 209.098  & - &450.43  & - \\
& PALOM \cite{lin_high-plex_2023} & 39.23$\pm$6.2212 & 50 & 250$\pm$ 101.3232 &50  \\
& ANTs \cite{avants_reproducible_2011} & 37.43$\pm$ 28.5700 & 90 & 170.27$\pm$ 78.5624 &90  \\
& Elastix \cite{klein2009elastix}  & 8.4304 $\pm$ 5.1031 & 35 & 145.69 $\pm$88.3913 &35  \\
& DFBR \cite{awan_deep_2023} & 2.8924 $\pm$  0.9012 & 50 & 5.392$\pm$1.6721 & 50.13 \\
& VALIS \cite{gatenbee_virtual_2023} & 2.2340$\pm$1.5612  & 60  & 2.5320 $\pm$ 1.7723 & 60  \\
& DeeperHistReg \cite{wodzinski_regwsi_2024} & 2.7072 $\pm$ 1.0001 & 70 & 3.6709 $\pm$ 1.3620 & 70 \\
% & VALIS} \cite{gatenbee_source_2023} &  &  & & \\
& \textbf{CORE} & \textbf{1.0272}{$\pm$0.3223} & \textbf{14.37} & \textbf{2.0495}{$\pm$0.6401} & \textbf{14.37} \\
\midrule
\textbf{Fine} & \textbf{CORE} & \textbf{0.36 }{$\pm$ 0.2210} & 1500  & \textbf{0.8460}{$\pm$0.5101} & 1500 \\
\bottomrule
\end{tabular}
\label{tab:reactivas}
\end{table}
\begin{figure}[H]  % Also force top placement
    \centering
    \includegraphics[width=1\textwidth]{figure_11.pdf}
    \captionsetup{width=\textwidth}
\caption{CORE results on the REACTIVAS dataset. WSI overview (0.625$\times$): low-magnification views of target (PAS), source (H\&E), and registered WSIs. Region overview (5$\times$): intermediate-resolution tissue regions across target, source, and registered images. High magnification (40$\times$): full-resolution local regions showing fine structural details. Registration overlays: comparison of target-source before registration, and target-registered alignments. Zoomed overlays (40$\times$): high-resolution visualization of registration quality at cellular scale.}
    \label{fig:result_reactivas}
\end{figure}
\begin{figure}[H]  % Also force top placement
    \centering
    \includegraphics[width=1\textwidth]{figure_12.pdf}
 \captionsetup{width=\textwidth}
\caption{Multi-resolution visualization of CORE results on the REACTIVAS dataset. (i) WSI overview (0.625$\times$): low-magnification views of target (mIF), source (H\&E), and registered WSIs. (ii) Region overview (5$\times$): intermediate-resolution tissue regions across target, source, and registered images. (iii) High magnification (40$\times$): full-resolution local regions showing fine structural details. (iv) Registration overlays: comparison of target-source before registration, and target-registered alignments. (v) Zoomed overlays (40$\times$): high-resolution visualization of registration quality at cellular scale.}
 \label{fig:result_mIF}
 \end{figure}

\subsection{Visualization Tool}
\begin{figure}[H]
    \centering
    \includegraphics[width=1.05\textwidth]{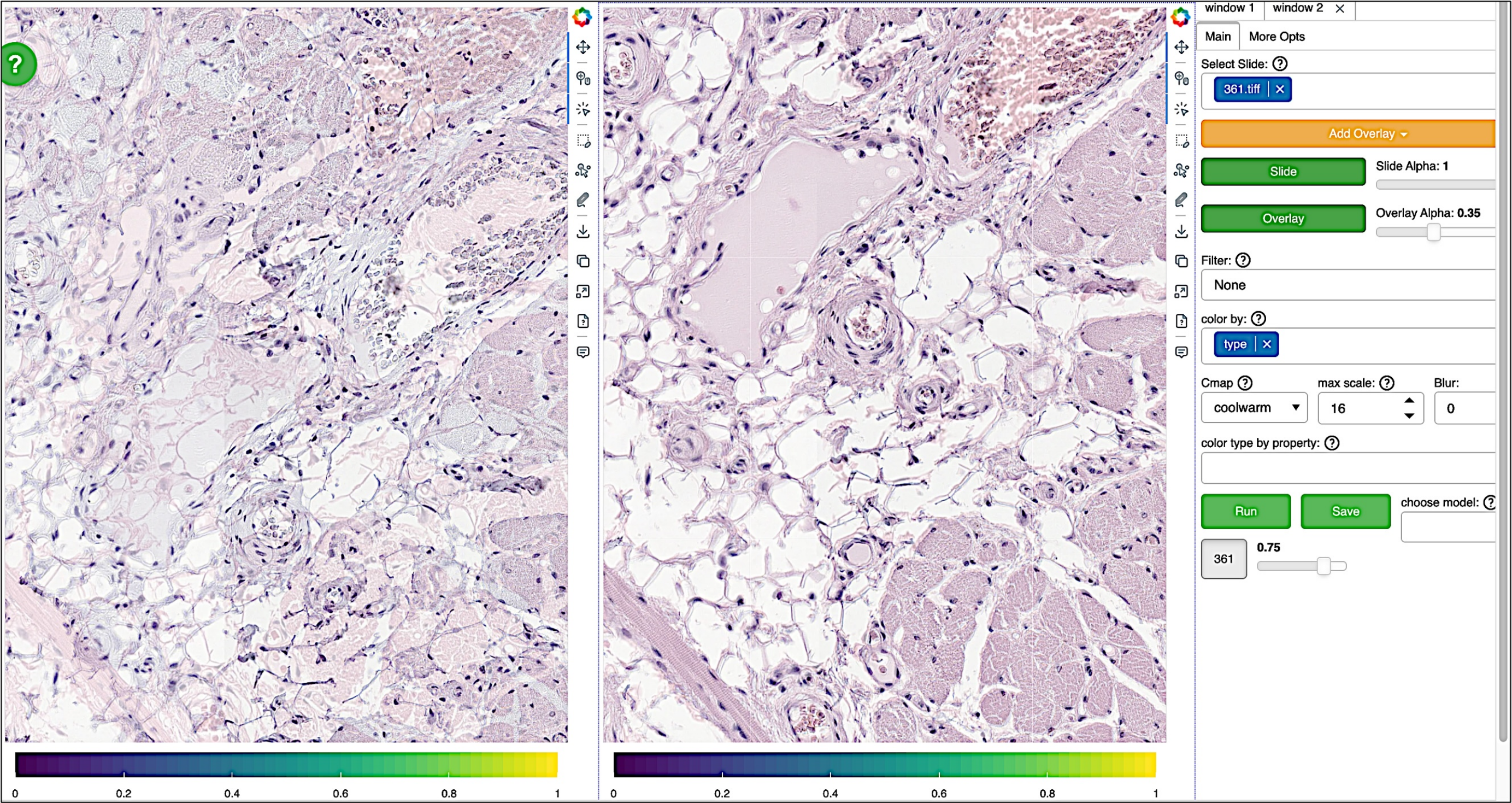}
    \captionsetup{width=\textwidth}
    \caption{TiaViz visualization tool displaying source (PHH3) and target (H\&E) re-stained slides before and after registration. (\textbf{Left}) Target overlaid on source prior to alignment. (\textbf{Right}) Target overlaid on source after alignment.}
    \label{fig:result_vis}
\end{figure}

A key step following registration is the visualization of the target and registered source images. However, saving WSIs after each registration step is both computationally and memory intensive, making it impractical for interactive analysis and iterative exploration. To address this limitation, TiaViz, applies spatial deformations on-the-fly, enabling users to visualize registered WSIs at any resolution without explicitly saving intermediate outputs. This design significantly reduces computational overhead and completes in seconds. Figure~\ref{fig:result_vis} illustrates TiaViz source(PHH3) and target (H\&E)  on the fly registered view, showing clear illustration of registration overlays after registration. Demo can be accessed at this link \href{https://tiademos.dcs.warwick.ac.uk/bokeh_app?demo=WSIReg}{CORE Demo} .

\subsection{Ablation Studies}

We conduct extensive ablation experiments to quantify the contribution of each component in CORE, including coarse registration design, feature extraction backbone, shape-aware nuclei matching, and point-set scale. These experiments isolate individual design choices to better understand their role in achieving robust multimodal WSI registration.

\subsubsection{Impact of Coarse Registration Components}

 Table~\ref{tab:ablation_rigid} evaluates the contribution of each component in the coarse registration stage, including morphology-only alignment (TriMorph), feature-only matching (XFeat), and the full CORE pipeline combining both. The results show that morphology-based alignment alone provides a strong structural prior but lacks robustness under large stain variation. In contrast, XFeat-based matching improves cross-modality correspondence by providing stain-invariant feature descriptors. However, the best performance is achieved when both components are integrated, confirming that geometric tissue priors and learned feature correspondences are complementary. Importantly, the combined CORE configuration reduces registration error without increasing runtime significantly, demonstrating that improved robustness does not come at the cost of computational efficiency.
\begin{table}[H]
\footnotesize
\centering
\captionsetup{width=\textwidth}
\caption{Ablation study on coarse registration components and feature extractors. Registration accuracy and runtime are compared using consecutive and re-stained (HyReCo) slides.}
\renewcommand{\arraystretch}{1.3}
\setlength{\tabcolsep}{6pt}

\begin{tabular}{p{3.5cm}
                p{1.6cm} p{1.6cm}
                p{1.6cm} p{1.6cm}}
\toprule

& \multicolumn{2}{c}{\textbf{Consecutive}} 
& \multicolumn{2}{c}{\textbf{Re-stained}} \\

\cmidrule(lr){2-3}
\cmidrule(lr){4-5}

\textbf{Rigid Method} 
& \textbf{AMTRE} & \textbf{Time (s)} 
& \textbf{AMTRE} & \textbf{Time (s)} \\

\midrule

TriMorph only                & 108.32 & 8 & 12.32  & 7 \\
XFeat only                   & 87.21 & 5 & 9.31 & 5 \\

\textbf{CORE}                & 65.29  & 10  &6.8 &10  \\

\bottomrule
\end{tabular}

\label{tab:ablation_rigid}
\end{table}

\subsubsection{Sensitivity to Shape-Aware Weighting}

We further investigated the influence of the shape-weight parameter $\delta$ in the proposed shape-aware distance metric (Table~\ref{tab:ablation_gamma_shape}). The parameter controls the balance between spatial proximity and morphological similarity during nuclei correspondence estimation. When $\delta = 0$, alignment relies purely on spatial distance, resulting in suboptimal registration accuracy. Introducing morphological constraints significantly improved performance, with the best overall results obtained around $\delta = 0.3$--$0.5$. Larger $\delta$ values overly emphasized shape similarity and degraded alignment robustness. These observations confirm that jointly modeling spatial and morphological information is essential for accurate cellular correspondence.

\begin{table}[H]
\footnotesize
\centering
\caption{Ablation study on the shape weight parameter ($\delta$) in the shape-aware distance metric. Registration accuracy is evaluated across different $\delta$ values on HyReCo, controlling the relative contribution of spatial proximity vs.\ morphological similarity during nuclei point-set alignment.}
\renewcommand{\arraystretch}{1.3}
\setlength{\tabcolsep}{4pt}
\begin{tabular}{p{3cm} p{1.5cm} p{1.5cm} p{1.5cm} p{1.5cm} p{1.5cm} p{1.5cm} p{1.5cm} p{1.5cm}}
\toprule
 & \multicolumn{2}{c}{\textbf{Consecutive}} & \multicolumn{2}{c}{\textbf{Re-stained}}  \\
\cmidrule(lr){2-3} \cmidrule(lr){4-5} 
\textbf{$\delta$} & \textbf{AMTRE} & \textbf{AATRE} & \textbf{AMTRE} & \textbf{AATRE}  \\
\midrule
0.0       &  65.21 & 109.81  & 5.61&  10.20  \\
0.05      & 60.32 &  107.36& 4.24 &   9.53 \\
0.1       & 54.32  & 94.23 & 3.82  & 8.21  \\
0.2       & 42.31 & 78.14 &  3.47 & 6.90  \\
\textbf{0.3 (CORE)} & 39.21 & 60.21 & \textbf{2.26}  & \textbf{5.41}  \\
0.4       & 40.23 & 65.37 & 3.1  &  6.22   \\
0.5       & \textbf{39.20} & \textbf{59.12}    & 2.54&  5.51  \\
0.7       & 56.12 & 98.44 &  4.5 &  9.23   \\
1.0       &  68.11 & 121.32 & 5.2  & 9.61   \\
\bottomrule
\end{tabular}
\label{tab:ablation_gamma_shape}
\end{table}

\subsubsection{Effect of Nuclei Point-Set Scale}

To analyze the scalability and robustness of the proposed shape-aware nuclei registration module, we varied the number of nuclei used during point-set matching (Table~\ref{tab:ablation_nuclei_density}). Increasing the nuclei count consistently improved registration accuracy, as reflected by lower AMTRE values. Although larger point sets increased computational cost, the proposed shape-aware registration remained computationally efficient even at high nuclei counts. In contrast, conventional CPD-based non-rigid registration exhibited rapidly increasing runtime complexity and became computationally impractical beyond 10k nuclei points. These findings demonstrate the efficient selection of sparse nuclei points enables computational time reduction in non linear registration stage.
\begin{table}[H]
\footnotesize
\centering
\caption{Ablation study on the impact of nuclei point-set number on registration accuracy and total registration time. AATRE and AMTRE are reported in $\mu$m. Time is in seconds.}
\renewcommand{\arraystretch}{1.3}
\setlength{\tabcolsep}{3pt}
\begin{tabular}{p{3cm}p{1.8cm}p{1.8cm}p{2.2cm}p{1.8cm}p{1.8cm}p{2.2cm}}
\toprule
\textbf{No.\ of Nuclei} &
\textbf{AMTRE [$\mu$m]} & \textbf{Shape-aware pointset Time (s)} & \textbf{AMTRE [$\mu$m]} & \textbf{CPD Non-rigid Time (s)} \\
\midrule
500          & 65.29 &   0.05 &   7.34 & 0.27 \\
1{,}000      & 57.23 &   0.05 &   4.96 & 1.45 \\
5{,}000      & 55.41 &   0.22 &   4.35 & 76.03 \\
10{,}000     & 53.12 &   0.49 &   4.35 & 481.81 \\
25{,}000     & 51.13 &   1.10 &  -- & -- \\
50{,}000     & 47.21 &   2.05 &  -- & -- \\
100{,}000    & 43.86 &   4.13 &  -- & -- \\
150{,}000    & 41.19 &   6.30 &  -- & -- \\
200{,}000    & 39.20 &   2.76 &  -- & -- \\
\bottomrule
\end{tabular}
\label{tab:ablation_nuclei_density}
\end{table}
\subsubsection{Per-Stain Robustness Analysis}

For evaluating the robustness of CORE for each stain, we performed extensive experiments separately for each stain in our datasets.
Tables~\ref{tab:ablation_per_stain_acrobat}, \ref{tab:ablation_per_stain_hyreco}, and \ref{tab:ablation_per_stain_comet} report performance across different staining protocols and datasets. Across all datasets, CORE maintains consistent performance despite substantial variations in staining appearance, imaging modality, and tissue morphology. Performance is generally stronger on re-stained and structurally consistent pairs, where morphological correspondence is preserved, and slightly weaker on highly heterogeneous immunohistochemical pairs with stronger structural divergence. These results confirm that CORE generalizes effectively across both bright-field and fluorescence-based modalities without dataset-specific tuning. Figure~\ref{fig:pointset_results} shows visualization of nuclei point sets registration on different tissue stains.

\begin{table}[H]
\footnotesize
\centering
\captionsetup{width=\textwidth}
\caption{Per-stain performance breakdown of CORE on the ACROBAT dataset.}
\renewcommand{\arraystretch}{1.3}
\setlength{\tabcolsep}{4pt}
\begin{tabular}{p{2cm} p{3cm} p{2cm} p{2cm} p{2cm} p{2cm}}
\toprule
\textbf{Dataset} & \textbf{Stain Pair} & \textbf{AATRE } & \textbf{AMTRE} & \textbf{AMxrTRE} & \textbf{No.\ Pairs} \\
\midrule
\multirow{5}{*}{ACROBAT}
& H\&E $\rightarrow$ ER    &  322.11 & 240.96 & 944.02 & 29 \\
& H\&E $\rightarrow$ Ki67  & 232.36  & 92.88 & 1133.95 & 29  \\
& H\&E $\rightarrow$ PGR   &  279.51 & 206.33 & 1215.56 & 28 \\
& H\&E $\rightarrow$ HER2  & 406.95 &  52.04 & 4161.93  & 14 \\

\bottomrule
\end{tabular}
\label{tab:ablation_per_stain_acrobat}
\end{table}

\begin{table}[H]
\footnotesize
\centering
\captionsetup{width=\textwidth}
\caption{Per-stain performance breakdown of CORE on the HyReCo dataset.}
\renewcommand{\arraystretch}{1.3}
\setlength{\tabcolsep}{4pt}
\begin{tabular}{p{2cm} p{3cm} p{2cm} p{2cm} p{2cm}}
\toprule
\textbf{Dataset} & \textbf{Stain Pair} & \textbf{AATRE } & \textbf{AMTRE} & \textbf{No.\ Pairs} \\
\midrule

% \midrule
\multirow{4}{*}{HyReCo}
& H\&E $\rightarrow$ PHH3  & 2.52 & 2.01  & 9  \\
& H\&E $\rightarrow$ CD8   & 6.86 & 5.92  & 9 \\
& H\&E $\rightarrow$ CD45  & 15.47 & 10.01  &9  \\
& H\&E $\rightarrow$ Ki67  & 5.83 & 4.78  &  9\\

\bottomrule
\end{tabular}
\label{tab:ablation_per_stain_hyreco}
\end{table}

\begin{table}[H]
\footnotesize
\centering
\captionsetup{width=\textwidth}
\caption{Per-stain performance breakdown of CORE on the Multi-IHC CRC dataset.}
\renewcommand{\arraystretch}{1.3}
\setlength{\tabcolsep}{4pt}
\begin{tabular}{p{2cm} p{3cm} p{2cm} p{2cm} p{2cm} p{2cm}}
\toprule
\textbf{Dataset} & \textbf{Stain Pair} & \textbf{AArTRE } & \textbf{AMTRE} & \textbf{AMxrTRE}  \\

\midrule
\multirow{3}{*}{Multi-IHC}
& H\&E $\rightarrow$ PMS2  & 0.00221 & 0.001560 & 0.00706\\
& H\&E $\rightarrow$ MSH6   & 0.00267& 0.001891 & 0.01304 \\
& H\&E $\rightarrow$ MSH2    & 0.00268 & 0.001231 & 0.00813   \\
& H\&E $\rightarrow$ MLH1    & 0.00373 & 0.00213 & 0.009210334 \\
\bottomrule
\end{tabular}
\label{tab:ablation_per_stain_comet}
\end{table}
\subsubsection{Per-Tissue Performance Analysis.}

We additionally evaluated registration performance across different tissue types in the ANHIR dataset (Table~\ref{tab:per_tissue_rtre}). The results indicate consistently low normalized rTRE values across multiple organs and histological structures, including breast, kidney, gastric, lung, and mammary tissues. While some variability exists due to differences in tissue morphology and section deformation, the framework maintained stable registration accuracy across all tissue categories. This demonstrates the robustness and generalizability of CORE for heterogeneous histopathology datasets.

\begin{table}[H]
\footnotesize
\centering
\captionsetup{width=\textwidth}
\caption{Per-tissue registration performance on the ANHIR dataset. rTRE values are normalised. }
\renewcommand{\arraystretch}{1.2}
\setlength{\tabcolsep}{4pt}
\begin{tabular}{lcccccc}
\toprule
\textbf{Tissue} & \textbf{AArTRE} & \textbf{AMrTRE} & \textbf{MArTRE} & \textbf{MMrTRE} & \textbf{AMxrTRE} & \textbf{MMxrTRE} \\
\midrule
COAD            & 0.0070 & 0.0054 & 0.0028 & 0.0017 & 0.0319 & 0.0234 \\
breast          & 0.0375 & 0.0381 & 0.0054 & 0.0036 & 0.0700 & 0.0232 \\
kidney          & 0.0141 & 0.0132 & 0.0037 & 0.0027 & 0.0292 & 0.0163 \\
gastric         & 0.0020 & 0.0014 & 0.0011 & 0.0007 & 0.0184 & 0.0089 \\
lung-lobes      & 0.0022 & 0.0010 & 0.0017 & 0.0009 & 0.0198 & 0.0166 \\
lung-lesion     & 0.0052 & 0.0032 & 0.0051 & 0.0033 & 0.0299 & 0.0242 \\
mice-kidney     & 0.0102 & 0.0089 & 0.0054 & 0.0017 & 0.0356 & 0.0240 \\
mammary-gland   & 0.0118 & 0.0092 & 0.0030 & 0.0014 & 0.0437 & 0.0202 \\
\midrule
\textbf{Overall} & 0.0040 & 0.00278 & 0.0026 & 0.0015 & 0.0323 & 0.0202 \\
\bottomrule
\end{tabular}
\label{tab:per_tissue_rtre}
\end{table}

Overall, the ablation studies demonstrate that the performance gains of CORE arise from the synergistic combination of fast feature-based coarse registration and shape-aware nuclei-level refinement. The experiments further show that integrating morphological priors during nuclei matching substantially improves cellular alignment accuracy while remaining computationally scalable for large WSIs.

Failure cases primarily occur in slides exhibiting severe tissue fragmentation, extremely sparse cellularity, substantial staining degradation, or extensive missing tissue regions where reliable structural correspondence cannot be established. In such cases, registration accuracy may degrade due to insufficient geometric constraints and unreliable nuclei correspondence. Although CORE demonstrates strong robustness across multiple staining modalities and tissue types, extremely large local deformation or highly degraded tissue morphology may still challenge accurate cell-level alignment.
\section{Discussion}
CORE combines global tissue-level alignment with local cell-centric refinement to address the inherent multi-scale deformation present in WSI registration. The central design principle is to progressively reduce the search space: from global structural alignment to fine-grained cellular correspondence. This staged formulation enables both robustness to large inter-slide variability and precision at the cellular scale, which is essential for downstream spatial analyses.
The CORE pipeline operates in two sequential stages. First, global coarse registration aligns tissue structures using low-resolution representations and robust feature matching. This step ensures that corresponding morphological regions lie within a reasonable spatial basin of convergence. Second, a cell-centric refinement stage models local non-rigid deformations using nuclei-level correspondences. This decomposition is critical because direct optimization at the cellular level without global alignment is highly unstable due to large initial mis-alignments in WSI pairs, where rotations, translations, and tissue cropping differences can exceed the capture range of local optimizers.

Since accurate tissue segmentation is a prerequisite for reliable global alignment. We evaluated two strategies:  a U-Net-based segmentation model, and a prompt-guided segmentation approach based on a foundation segmentation model. U-Net improves segmentation consistency but introduces two limitations: increased computational cost and reduced generalization across unseen staining domains unless retrained. In contrast, the prompt-guided segmentation approach provides consistent tissue masks across diverse modalities without dataset-specific training. This robustness is particularly important in WSI registration, where stain variability and scanner differences can significantly affect intensity distributions. In practice, the quality of segmentation directly influences downstream feature stability and therefore registration accuracy.

For coarse registration, we evaluated both classical and deep learning-based feature extractors for correspondence estimation under severe appearance variation. Classical descriptors such as  SIFT \cite{lindeberg_scale_2012}, SURF \cite{leonardis_surf_2006}, ORB \cite{rublee_orb_2011}, HARRIS \cite{yi-bo_harris_2011}, BRIEF \cite{hutchison_brief_2010}, and FAST \cite{guo_novel_2011} show limited robustness in histopathology due to their reliance on local intensity gradients, which are highly stain-dependent. As a result, they frequently produce inconsistent or incorrect correspondences under cross-stain conditions. Recent deep matching methods, including LoFTR \cite{sun2021loftr} and RoMa \cite{edstedt2024roma}, improve robustness in natural image settings but exhibit a domain gap when applied to histopathology data, where textures are repetitive, weakly structured, and stain-dependent. In this context, XFeat provides a more stable trade-off between discriminability and generalization. Its multi-scale representation improves robustness to staining variation while maintaining computational efficiency. Importantly, the reduction in spurious correspondences improves the stability of downstream geometric estimation, preventing degenerate transformations during coarse registration. Figure~\ref{fig:acrobat_reg_artifacts} illustrates CORE registration results after coarse registration in the presence of artifacts (e.g., tissue folds) and demonstrates the robustness of the proposed approach.

We observe a clear trade-off between spatial resolution and registration performance at coarse level registration. At very low resolutions, fine structural details are lost, reducing the discriminability of features and leading to ambiguous correspondences. At higher resolutions, although feature richness improves, computational cost increases significantly without proportional gains in registration accuracy. Empirically, intermediate resolutions (0.625$\times$ to 1.25$\times$) provide the best balance between structural fidelity and efficiency. At this scale, global tissue morphology is preserved while suppressing high-frequency noise that can interfere with robust matching. 

We also find that applying point-set registration without prior coarse alignment leads to unstable optimization. In such cases, the initial displacement between corresponding nuclei lies outside the convergence basin of standard point-set objectives, resulting in incorrect correspondences and distorted transformations. This confirms that coarse alignment is not optional but structurally necessary. It ensures that corresponding tissue regions are approximately aligned, allowing the subsequent nuclei-based refinement to focus on modeling local non-rigid deformations rather than recovering global pose differences.
Using coarse alignment as initialization, we compare state-of-the-art point-set registration methods for local refinement. ICP \cite{zhang_iterative_1994} is fast but sensitive to outliers and repetitive structures, which are common in histopathology. Gaussian Mixture Model (GMM) \cite{eckart_fast_2018} based methods provide improved robustness but incur higher computational cost, particularly at large point-set scales. Tree-accelerated variants improve efficiency but still depend strongly on initialization quality. Across all methods, performance degrades significantly without reliable coarse alignment, reinforcing the importance of a hierarchical registration strategy. In contrast, the proposed cell-centric refinement explicitly leverages nuclei structure as biologically meaningful anchors, enabling more stable correspondence estimation in dense and repetitive tissue environments.

For nuclei detection, we adopt a classical image-processing pipeline to maximize robustness across datasets. Unlike deep learning-based methods, which require large annotated datasets and often suffer from reduced generalization across staining protocols, tissue types, and imaging systems, classical approaches based on stain decomposition and morphological filtering provide reliable nuclei localization without any training data. Although this may result in lower detection precision, it is sufficient for CORE, which operates on sparse nuclei representations rather than exhaustive nuclei detection. This design is further motivated by the computational complexity of coherent point drift (CPD), which increases with the number of detected nuclei. Prior work has shown that sparse nuclei representations can achieve registration accuracy comparable to dense approaches while substantially improving computational efficiency~\cite{yap_nuclei-location_2024}. Consequently, CORE focuses on extracting representative nuclei centroids that capture the underlying spatial tissue structure, avoiding the additional computational burden associated with dense segmentation of overlapping nuclei while maintaining effective registration performance.

\paragraph{Computational complexity}
CORE is designed to scale to gigapixel WSIs by progressively constraining the optimization space. Let $k$ denote the number of extracted keypoints and $N, M$ the number of nuclei in source and target images. Coarse feature matching scales approximately as $O(k \log k)$ using nearest-neighbor search structures, though this depends on feature dimensionality and distribution. 
The point-set refinement stage based on probabilistic correspondence estimation scales approximately as $O(TNM)$, where $T$ is the number of optimization iterations. Table~\ref{tab:gpu_cpu_point} shows per step computation time on CPU and GPU for our CORE algorithm.
To maintain tractability, we employ sparse nuclei sampling and coarse-to-fine initialization, ensuring that the refinement stage operates on reduced and well-aligned point sets. In practice, coarse registration completes in a few seconds, while fine registration requires takes  few minutes depending on tissue size and nuclei density, reflecting the trade-off between accuracy and computational cost.

% Our experiments highlight three key observations. First, prompt-guided tissue segmentation provides robust and generalizable tissue masks across heterogeneous staining conditions. Second, modern learned feature representations such as XFeat outperform classical descriptors and general-purpose deep matchers in histopathology due to better robustness to repetitive textures and stain variability. Third, intermediate-resolution coarse registration provides the optimal trade-off between computational efficiency and structural fidelity.
% Importantly, the proposed hierarchical design enables coarse alignment to place corresponding structures within a stable convergence basin, while the cell-centric refinement stage achieves fine-grained alignment by leveraging nuclei-level structure. This synergy is essential for achieving sub-cellular registration accuracy across diverse staining modalities. 
Despite strong performance, CORE has some limitations. The most computationally expensive component is the nuclei-based refinement stage, which limits scalability to extremely large datasets without parallelization. Additionally, while the framework demonstrates robustness across a wide range of staining protocols, performance under rare or highly atypical staining conditions has not been exhaustively evaluated. Future work may explore learning-based acceleration of nuclei detection and correspondence estimation to reduce runtime while preserving generalization. Self-supervised or domain-adaptive feature learning tailored to histopathology may further improve robustness under extreme stain variation. Another promising direction is integrating uncertainty estimation into the registration pipeline to better handle ambiguous correspondences in highly heterogeneous tissue regions. In summary, this study demonstrates that combining deep feature-based global alignment with biologically informed local refinement is a powerful strategy for multimodal WSI registration. Our findings provide a foundation for the advancement of spatial analysis in digital pathology and the integration of complex, multidimensional biological data. 

\section{Conclusions}

In this work, we introduced \textbf{CORE}, a coarse-to-fine registration framework for WSI alignment across diverse histopathological staining modalities, including H\&E, mIHC, PAS, CyC-IF and mIF. The proposed framework addresses the challenges of multimodal WSI registration by integrating efficient preprocessing, robust feature matching, and biologically informed local refinement within a unified pipeline. CORE begins with tissue region extraction using prompt-guided Florence-SAM segmentation, allowing accurate and reliable isolation of tissue structures with varying stain appearances. An initial rigid alignment is then estimated from the geometric properties of the extracted tissue masks, providing a stable global initialization. To further refine the alignment, XFeat is employed for dense deep feature extraction and matching, followed by iterative optimization of coarse deformation field with adaptive regularization to preserve structural consistency while accommodating tissue deformation. To achieve high-precision local alignment, the nuclei are detected from both the target slide and the coarsely registered source slide at full-resolution. Subsequently, a shape-aware point-set registration strategy is applied, followed by deformable refinement using the CPD algorithm, enabling nuclei-level registration accuracy. Extensive evaluation across multiple datasets demonstrates that CORE achieves superior registration performance while maintaining strong computational efficiency, consistently outperforming existing approaches in terms of alignment accuracy. By combining coarse tissue-level alignment with fine-grained cell-centric refinement, the proposed framework provides a robust and annotation-free solution for multimodal histopathological registration. This capability is particularly important for downstream applications that require precise cross-modality correspondence, including spatial transcriptomics, single-cell analysis, and computer-aided diagnostic systems.

\section{Code Availability}

To facilitate reproducibility and encourage further research, we release the complete implementation of CORE as open source. The full source code is publicly available on GitHub\footnote{\url{https://github.com/eshasadia/WSI\_mIF\_Reg.git}}. In addition, we provide an interactive web-based demonstration through TiaViz\footnote{\url{https://tiademos.dcs.warwick.ac.uk/bokeh\_app?demo=WSIReg}}, allowing users to explore registration and deformation results at full whole-slide image resolution.

\section{Acknowledgments}
SEAR and BE report financial support by the MRC (MR/X011585/1). SEAR and NR are grateful to the European Commission, which provided funding through the Innovative Medicines Initiative 2 Joint Undertaking under grant agreement No 945358. AS and SEAR received financial support from the British Council (project ref: 1203764217). The REACTIVAS study was partly supported from an Investigator-Initiated Program of Merck Sharp \& Dohme Corp (MSD). The opinions expressed are those of the authors and do not necessarily represent those of MSD. NR is CEO and CSO of Histofy Ltd.

\bibliographystyle{unsrt}
\bibliography{references.bib}

\clearpage
\appendix

\setcounter{table}{0}
\setcounter{figure}{0}
\renewcommand{\thetable}{A.\arabic{table}}
\renewcommand{\thefigure}{A.\arabic{figure}}

\section{Supplementary Information}

\paragraph{Using the demo:}
\label{app:tiaviz}

To facilitate real-time deformation of the source image and simultaneous visualization of the transformed and target WSIs, we provide a dedicated viewer, \textit{TIAViz} \cite{eastwood_tiaviz_2024}. The tool is available as part of the \textit{TIATOOLBOX} library \cite{pocock_tiatoolbox_2022}, which can be installed via \texttt{pip}. The demo can be used by following the steps below:

\begin{enumerate}
    \item \textbf{Install TIAViz.}  
    Install the \textit{TIATOOLBOX} package using \texttt{pip}.

    \item \textbf{Prepare the directory structure.}  
    Within the working directory, create two subdirectories named \texttt{slides} and \texttt{overlays}.

    \item \textbf{Place the required files.}  
    Store the source WSI in the \texttt{slides} directory.  
    Store the target WSI together with  deformation field in the \texttt{overlays} directory.

    \item \textbf{Ensure filename consistency.}  
    All corresponding files must share the same stem name so that TIAViz can automatically match the source, target, and deformation field.

    \item \textbf{Launch the viewer.}  
    Start the TIAViz interface using the \texttt{tiavisualize} command, either from the command line or within a Python notebook, specifying the paths to the \texttt{slides} and \texttt{overlays} directories. 

\item \textbf{Registration visualization.}  
From the \texttt{slides} drop-down in the top-left panel, first select the source slide. After this, choose the deformation field from the \texttt{overlay} drop-down; the deformation will be applied to the source slide within a few seconds. Next, select the target slide from the \texttt{Overlay} drop-down to display it on top of the registered source. Use the right-hand panel to adjust the alpha values of both layers to achieve optimal visualization of the alignment.

\end{enumerate}

\paragraph{Stain normalization:}
\label{app:stain}
Stain normalization is often required for H\&E slides because their color appearance varies widely across laboratories, scanners, and staining batches, which can affect how prompt-based models such as Florence-SAM \cite{xiao_florence-2_2023} detect tissue boundaries. Normalizing H\&E reduces this variability and provides a more consistent input for tissue mask extraction. In contrast, IHC slides uses stable chromogen such as DAB that have less color variability, and convey biologically meaningful intensity differences that normalization could distort, so stain normalization is generally unnecessary for IHC. Importantly, CORE does not rely on strict stain consistency. Instead, the framework primarily exploits structural and morphological information, enabling robust alignment even when stain normalization is imperfect or unavailable.

\section{Supplementary Tables}
\renewcommand{\thetable}{B.\arabic{table}}
\renewcommand{\thefigure}{B.\arabic{figure}}

\begin{table}[H]
\centering
\footnotesize
\captionsetup{width=\textwidth}
\caption{{Overview of stain types, biomarkers, and imaging modalities across the WSI registration datasets. The table highlights the diversity of histopathological stains and acquisition modalities, including H\&E, IHC, and mIF imaging, demonstrating the robustness and generalizability requirements of the proposed registration framework across multi-stain and multi-modal scenarios.}}
\label{tab:stains}

\renewcommand{\arraystretch}{1.2}
\setlength{\tabcolsep}{5pt}

\begin{tabular}{p{2.8cm} p{2.3cm} p{7.5cm} p{2.5cm}}
\toprule

\textbf{Dataset} &
\textbf{Marker} &
\textbf{Stain Name} &
\textbf{Modality} \\

\midrule

\multirow{5}{*}{\textbf{ACROBAT}}
& H\&E & Hematoxylin and Eosin & Bright-field \\
& ER & Estrogen Receptor & IHC \\
& PR & Progesterone Receptor & IHC \\
& HER2 & Human Epidermal Growth Factor Receptor 2 & IHC \\
& Ki-67 & Marker of Proliferation Ki-67 & IHC \\

\midrule

\multirow{5}{*}{\textbf{HyReCo}}
& H\&E & Hematoxylin and Eosin & Bright-field \\
& CD8 & Cluster of Differentiation 8 (Cytotoxic T cells) & IHC \\
& CD45 & Leukocyte Common Antigen Isoforms & IHC \\
& Ki-67 & Marker of Proliferation Ki-67 & IHC \\
& PHH3 & Phospho-Histone H3 & IHC \\

\midrule

\multirow{17}{*}{\textbf{ANHIR}}
& H\&E & Hematoxylin and Eosin & Bright-field \\
& CD1a & Cluster of Differentiation 1a & IHC \\
& CD31 & Platelet Endothelial Cell Adhesion Molecule-1 & IHC \\
& CD4 & Cluster of Differentiation 4 & IHC \\
& CD68 & Macrophage Marker CD68 & IHC \\
& CD8 & Cluster of Differentiation 8 & IHC \\
& Cc10 & Club Cell Secretory Protein (CC10) & IHC \\
& EBV & Epstein–Barr Virus Marker & IHC \\
& ER & Estrogen Receptor & IHC \\
& HER2 & Human Epidermal Growth Factor Receptor 2 & IHC \\
& Ki67 & Marker of Proliferation Ki-67 & IHC \\
& MAS & Macrophage Activation System Marker & IHC \\
& PAS & Periodic Acid–Schiff Stain & Bright-field \\
& PASM & Periodic Acid–Silver Methenamine Stain & Bright-field \\
& PR & Progesterone Receptor & IHC \\
& Pro-SPC & Surfactant Protein C Precursor & IHC \\
& aSMA & Alpha-Smooth Muscle Actin & IHC \\

\midrule

\multirow{3}{*}{\textbf{REACTIVAS}}
& PAS & Periodic Acid–Schiff Stain & Bright-field \\
& H\&E & Hematoxylin and Eosin & Bright-field \\
& mIF & Multiplex Immunofluorescence & Fluorescence \\

\midrule

\multirow{6}{*}{\textbf{Multi-IHC CRC}}
& H\&E & Hematoxylin and Eosin & Bright-field \\
& CK818 & Cytokeratin 8/18 & IHC \\
& MLH1 & MutL Homolog 1 & IHC \\
& MSH2 & MutS Homolog 2 & IHC \\
& MSH6 & MutS Homolog 6 & IHC \\
& PMS2 & Postmeiotic Segregation Increased 2 & IHC \\

\bottomrule
\end{tabular}
\end{table}
\begin{table}[H]

\centering
\caption{Configuration parameters used for coarse-level deformation estimation during the non-rigid registration stage. The table summarizes the selected similarity metric, optimization strategy, regularization approach, interpolation method, learning rate, and iteration range.}
\label{tab:registration_params}
\setlength{\tabcolsep}{3pt} % adjust spacing as needed
% alternating rows starting from 3rd
\begin{tabular}{@{}p{5cm} p{3.5cm}@{}}
\toprule
\textbf{Parameter} & \textbf{Value} \\
\midrule
Similarity Metric       & NCC \\
optimizer               & Adam \\
Regularizer             & Adaptive \\
Interpolation Method    & Bilinear \\
Learning Rate           & 0.001 \\
Iterations              & 200$\sim$500 \\
\bottomrule
\end{tabular}
\label{tab:nonrigid}
\end{table}

\begin{table}[H]

\centering
\captionsetup{width=\textwidth}
\caption{Parameter settings used in the proposed shape-aware nuclei-based registration framework. The table summarizes the key optimization and deformation control parameters, including their default values, and allowable ranges. These parameters govern the balance between spatial and morphological alignment, optimization convergence, regularization strength, and deformation smoothness for robust nuclei-level registration.}
\setlength{\tabcolsep}{1pt} % Adjust column separation % Alternating row colors, starting from 3rd row
\begin{tabular}{p{3cm}p{2cm}p{6cm}p{3cm}p{2cm}}
\toprule
\textbf{Parameter} & \textbf{Symbol} & \textbf{Description} & \textbf{Default} & \textbf{Range} \\
\midrule
Shape Weight & $\delta$ & Spatial vs. shape trade-off & 0.3 & [0.0, 1.0] \\
Tolerance & $\varepsilon$ & optimization convergence threshold & $10^{-8}$ & [$10^{-8}$, $10^{-4}$] \\
Max Iterations & $i_{\max}$ & Maximum number of optimization steps & 100 & [50, 500] \\
Reg. Strength & $\lambda$ & Weight of regularization penalty & 1.0 & [0.1, 10.0] \\
Smoothness & $\beta$ & Controls smoothness of deformation field & 2.0 & [1.0, 5.0] \\
k-d Leaf Size & $k$ & Leaf size for nearest-neighbor queries & 10 & [5, 50] \\
\arrayrulecolor{black}\bottomrule
\end{tabular}
\label{tab:algoparams}
\end{table}

\begin{table}[H]

\centering
\captionsetup{width=\textwidth}
\caption{Parameter configuration for the non-rigid registration stage of the proposed framework. The table presents the key parameters. These parameters are designed to ensure stable convergence, smooth deformation modeling, and accurate alignment across fine nonrigid registration stage.}
\setlength{\tabcolsep}{1pt} % adjust spacing as needed
 % alternating rows starting from 3rd
\begin{tabular}{@{}p{3cm} p{2cm} p{2cm} p{4cm} p{3.5cm}@{}}
\toprule
\textbf{Parameter} & \textbf{Symbol} & \textbf{Default} & \textbf{Range} & \textbf{Description} \\
\midrule
regularization Weight & $\alpha$ & 0.01 & [0.001, 0.1] & Controls smoothness vs. data fidelity \\
Smoothness Parameter & $\beta$ & 0.5 & [0.1, 2.0] & Gaussian kernel bandwidth \\
Maximum Iterations & $i_{\max}$ & 200 & [100, 500] & EM algorithm iteration limit \\
Convergence Tolerance & $\varepsilon$ & $1 \times 10^{-9}$ & [$1 \times 10^{-12}$, $1 \times 10^{-6}$] & Weight change threshold \\
Outlier Weight & $w$ & 0.1 & [0.01, 0.3] & Uniform distribution weight \\
Initial Variance & $\sigma^2_{\text{init}}$ & 1.0 & [0.1, 5.0] & Initial Gaussian variance \\
Gaussian Smoothing & $\sigma$ & 10.0 & [5.0, 20.0] & Dense field smoothing \\
Max Displacement & $d_{\max}$ & 10.0 & [5.0, 50.0] & Maximum pixel displacement \\
\bottomrule
\end{tabular}
\label{tab:cpd}
\end{table}

\begin{table}[H]
\footnotesize
\centering
\captionsetup{width=\textwidth}
\caption{Definitions of relative Target Registration Error (rTRE) metrics used in evaluation.}
\begin{tabular}{p{2cm} p{6cm}}
\toprule
\textbf{Abbreviation} & \textbf{Definition} \\
\midrule
AArTRE   & Average Absolute relative Target Registration Error \\
AMrTRE   & Average Median relative Target Registration Error \\
MArTRE   & Median Absolute relative Target Registration Error \\
MMrTRE   & Median Median relative Target Registration Error \\
AMxrTRE  & Average Maximum relative Target Registration Error \\

\bottomrule
\end{tabular}
\label{tab:abb}
\end{table}
\begin{table}[H]
\footnotesize
\centering
\captionsetup{width=\textwidth}
\caption{Performance comparison of \textbf{CORE} rigid  registration method on the \textbf{HyReCo} Dataset (Consecutive and re-stained tissue sections).}
\renewcommand{\arraystretch}{1.2}
\begin{tabular}{
p{1.4cm}  % New column for Rigid/Non-rigid
p{1.5cm}  % Resolution (for Rigid rows)
p{2.5cm}  % Technique
p{1.4cm}  % Consecutive AMTRE
p{1.4cm}  % Consecutive AATRE
p{1.4cm}  % Consecutive Time
p{1.4cm}  % re-stained AMTRE
p{1.4cm}  % re-stained Time
}
\toprule
\multirow{2}{*}{\textbf{Type}} & \multirow{2}{*}{\textbf{Resolution}} & \multirow{2}{*}{\textbf{Technique}} &
\multicolumn{3}{c}{\textbf{Consecutive Sections}} &
\multicolumn{2}{c}{\textbf{Re-stained Sections}} \\
\cmidrule(lr){4-6} \cmidrule(lr){7-8}
& & & \textbf{AMTRE [$\mu$m]} & \textbf{AATRE [$\mu$m]} & \textbf{Avg Time (s)} &
\textbf{AMTRE [$\mu$m]} & \textbf{Avg Time (s)} \\
\midrule
% --- RIGID SECTION ---
\multirow{7}{*}{\textbf{Rigid}} & \multirow{5}{*}{\textbf{Coarse}} 
& DFBR \cite{awan_deep_2023} & 542.42 & 1754.52 & 240 & 30.03 & 50 \\
&& AGH \cite{wodzinski_multistep_2020} & 442.73 & 753.13 & 150 & 28.58 & 120 \\
&& \textbf{KAZE} & - & - & &27.2317 & 35.621  \\
&& DeeperHistReg \cite{wodzinski_regwsi_2024} & 174.9 & 234.5 & 120 & 12.24 & 70 \\
&& \textbf{SURF} & 158.93  & 187.144  &  42.866  &  7.1134  &  42.866  \\
&& \textbf{ \cite{lindeberg_scale_2012}} & 149.8 &168.65 & 30.145 & 13.2379  &  30.145 \\
&& \textbf{BRISK \cite{leutenegger_brisk_2011}} &  149.89 &   150.53 &  40 &   -&  - \\
&& \textbf{MSER \cite{matas2004robust}} &  143.12 &  1453.04 &  49.229 &  7.1641 &  49.229\\
&& \textbf{HARRIS \cite{harris1988combined}} &  134.0032  &  142.9819 & - & -  & - \\
&& HistokatFusion \cite{lotz_comparison_2023} & \textbf{20.12} & \textbf{62.58} & 50.13 & \textbf{2.00} & 50.14 \\
&& \textbf{CORE} & 65.26 & 109.83 & \textbf{18} & 6.81 & \textbf{18} \\
\midrule
& \textbf{Fine} & \textbf{CORE} & 39.20 & \textbf{60.21} & 2340  & \textbf{2.26} & 2340 \\
\bottomrule
\end{tabular}
\label{tab:hyreco-rigid}
\end{table}

\begin{table}[H]
\footnotesize
\centering
\captionsetup{width=\textwidth}
\caption{Stepwise results on the \textbf{ANHIR} dataset using average aggregation. Each alignment step progressively reduces error Median aggregation results.}
\renewcommand{\arraystretch}{1.2}
\setlength{\tabcolsep}{2pt}
\begin{tabular}{lccc}
\toprule
\textbf{Step} & \textbf{AArTRE} & \textbf{AMrTRE} & \textbf{AMxrTRE} \\
\midrule
Initial             & 0.1340 &0.0665  & 0.2338 \\
Coarse Transform     & 0.0040  & 0.0015 & 0.0240  \\
Fine  Transform & 0.0034  & 0.0012  & 0.0323    \\
\hline
\end{tabular}
\label{tab:anhir_step}
\end{table}

\begin{table}[H]
\footnotesize
\centering
\captionsetup{width=\textwidth}
\caption{Stepwise results on the \textbf{HyReCo} consecutive tissue sections using average aggregation. Each alignment step progressively reduces error Median aggregation results.}
\renewcommand{\arraystretch}{1.2}
\setlength{\tabcolsep}{2pt}
\begin{tabular}{lccc}
\toprule
\textbf{Step} & \textbf{AATRE} & \textbf{AMTRE} & \textbf{AMxTRE} \\
\midrule
Initial             & 3082.35 & 2583.96 & 7519.2 \\
Coarse Transform     & 105.3  &65.5   & 253.1  \\
Fine Transform    & 5.02   & 4.35   & 50.83    \\
\hline
\end{tabular}
\label{tab:hyreco_step}
\end{table}

\begin{table}[H]
\footnotesize
\centering
\captionsetup{width=\textwidth}
\caption{Stepwise results on the \textbf{Multi-IHC CRC} using average aggregation. Each alignment step progressively reduces error Median aggregation results.}
\renewcommand{\arraystretch}{1.2}
\setlength{\tabcolsep}{2pt}
\begin{tabular}{lccc}
\hline
\textbf{Step} & \textbf{AArTRE} & \textbf{AMrTRE} & \textbf{AMxrTRE} \\
\toprule
Initial             &  0.18079 & 0.15209 & 0.36154  \\
Coarse Transform     &  0.00431 &  0.0024&   0.0153\\
Fine  Transform &  0.002281  & 0.0011   &  0.0142  \\
\midrule
\end{tabular}
\label{tab:comet_step}
\end{table}

\begin{table}[H]
\footnotesize
\centering
\captionsetup{width=\textwidth}
\caption{Stepwise results on the \textbf{REACTIVAS} using average aggregation. Each alignment step progressively reduces error Median aggregation results.}
\renewcommand{\arraystretch}{1.2}
\setlength{\tabcolsep}{2pt}
\begin{tabular}{lccc}
\hline
\textbf{Step} & \textbf{AATRE} & \textbf{AMTRE} & \textbf{AMxTRE} \\
\hline
Initial             & 209.098  &  150.43 &400.51  \\
Coarse Transform     &  1.57 &1.0272   & 5.76  \\
Fine Transform & 0.45   &   0.36  &  1.24  \\

\hline
\end{tabular}
\label{tab:reactivas_step}
\end{table}

\begin{table}[H]
\footnotesize
\centering
\captionsetup{width=\textwidth}
\caption{Comparison of average processing times for each stage of the proposed registration pipeline on CPU and GPU platforms. The table highlights the computational efficiency of different preprocessing, feature extraction, similarity estimation, and registration components, demonstrating the acceleration achieved through GPU-based execution for large-scale WSI registration tasks.}
\label{tab:step_timing}
\setlength{\tabcolsep}{4pt}
\begin{tabular}{@{}p{6cm} p{5cm} p{5cm}@{}}
\toprule
\textbf{Step} & \textbf{Average Time CPU (sec)} & \textbf{Average Time GPU (sec)} \\
\midrule
Initial                         & 0.29          & 0.50 \\
Downsampling                   & 0.80          & 0.67 \\
Preprocessing                  & 0.19          & 0.00 \\
Tissue Mask Extraction         & 6.08          & 5.40 \\
TriMorph                       & 1.22          & 0.56 \\
Evaluation                     & 0.70          & 0.29 \\
XFeat                          & 2.91          & 0.80 \\
Similarity Estimation          & 0.73          & 0.29 \\
Warping                        & $<$0.01       & 0.00 \\
Fine Point-set Rigid           & 456            & 299 \\
Fine Point-set Non-Rigid       & 400  & 400 \\
\bottomrule
\end{tabular}
\label{tab:gpu_cpu_point}
\end{table}

\section{Supplementary Figures}
\renewcommand{\thetable}{C.\arabic{table}}
\renewcommand{\thefigure}{C.\arabic{figure}}
Figure~\ref{fig:staining} shows comparison of tissue staining techniques: (Left) Hematoxylin staining highlights cell nuclei in blue. (Middle) Immunohistochemistry using an antibody-chromogen system detects a specific protein in brown. (Right) Multiplex immunofluorescence allows simultaneous visualization of multiple markers, with nuclei staining with DAPI coloured in blue.
\begin{figure}[H]  % Also force top placement
    \centering
    \includegraphics[width=0.8\textwidth]{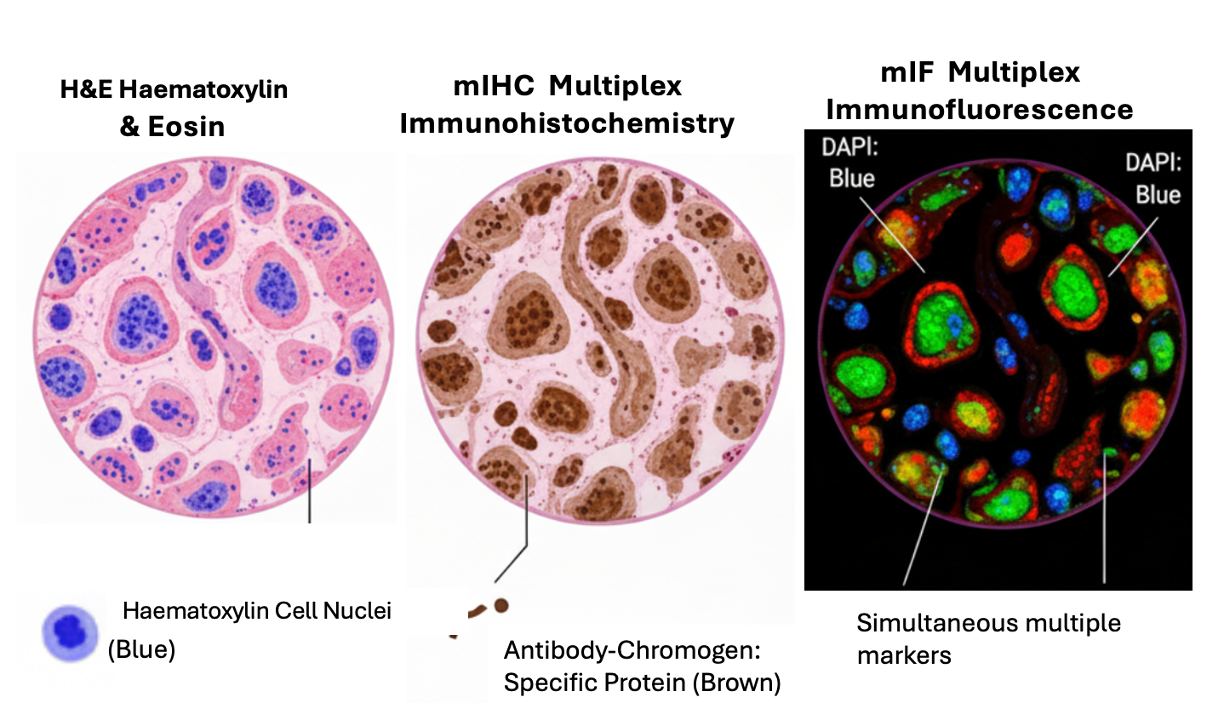}
    \caption{ Representative examples of different histopathological staining modalities used in this study: (1) H\&E, (2) mIHC, and (3) mIF. These staining variations highlight the significant appearance and modality differences encountered during WSI registration.}
    \label{fig:staining}
\end{figure}
\begin{figure}[H]  % Also force top placement
    \centering
    \includegraphics[width=\textwidth]{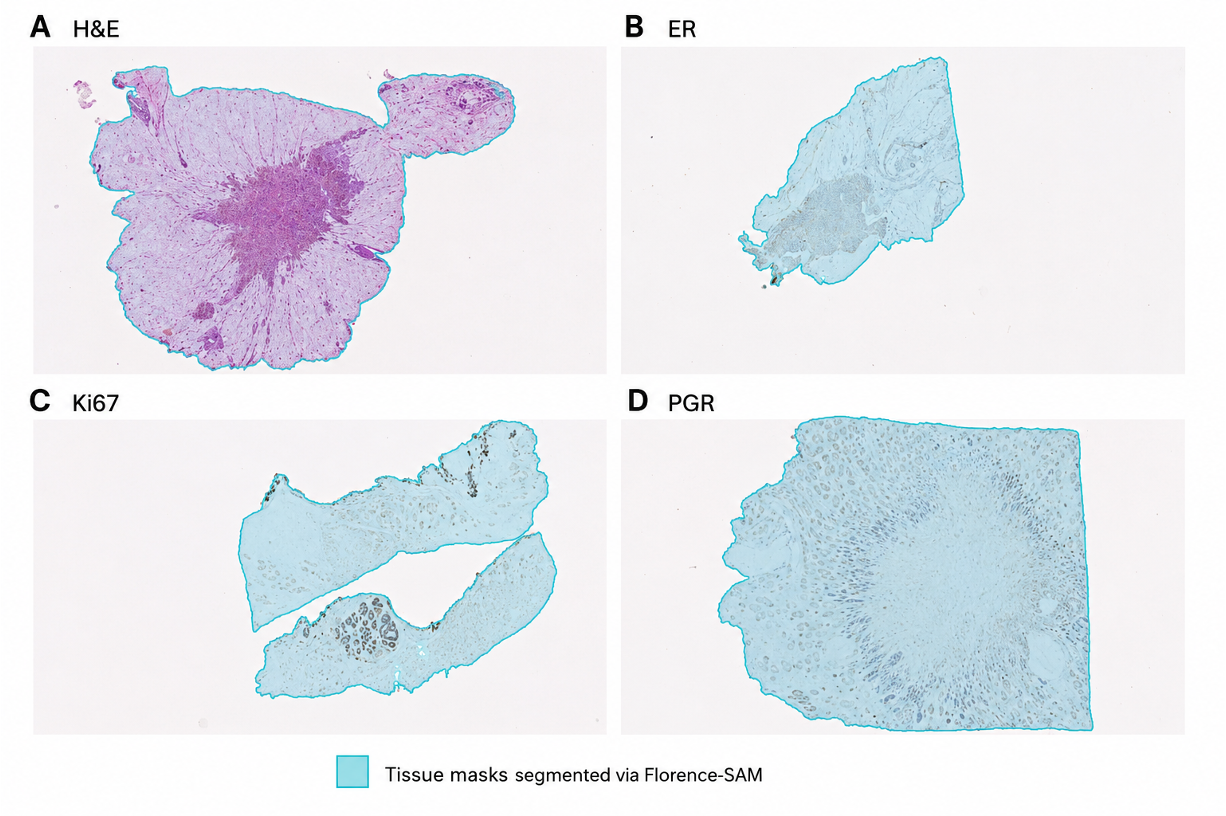}
    \captionsetup{width=\textwidth}
  \caption{Examples of tissue masks generated using the proposed prompt-guided Florence-SAM tissue segmentation framework. The generated masks accurately identify tissue regions across diverse staining modalities and complex histopathological structures, providing robust preprocessing support for subsequent registration tasks.}
 \label{fig:tissue_masks}
\end{figure}
\begin{figure}[H]
    \centering
    \includegraphics[width=1\textwidth]{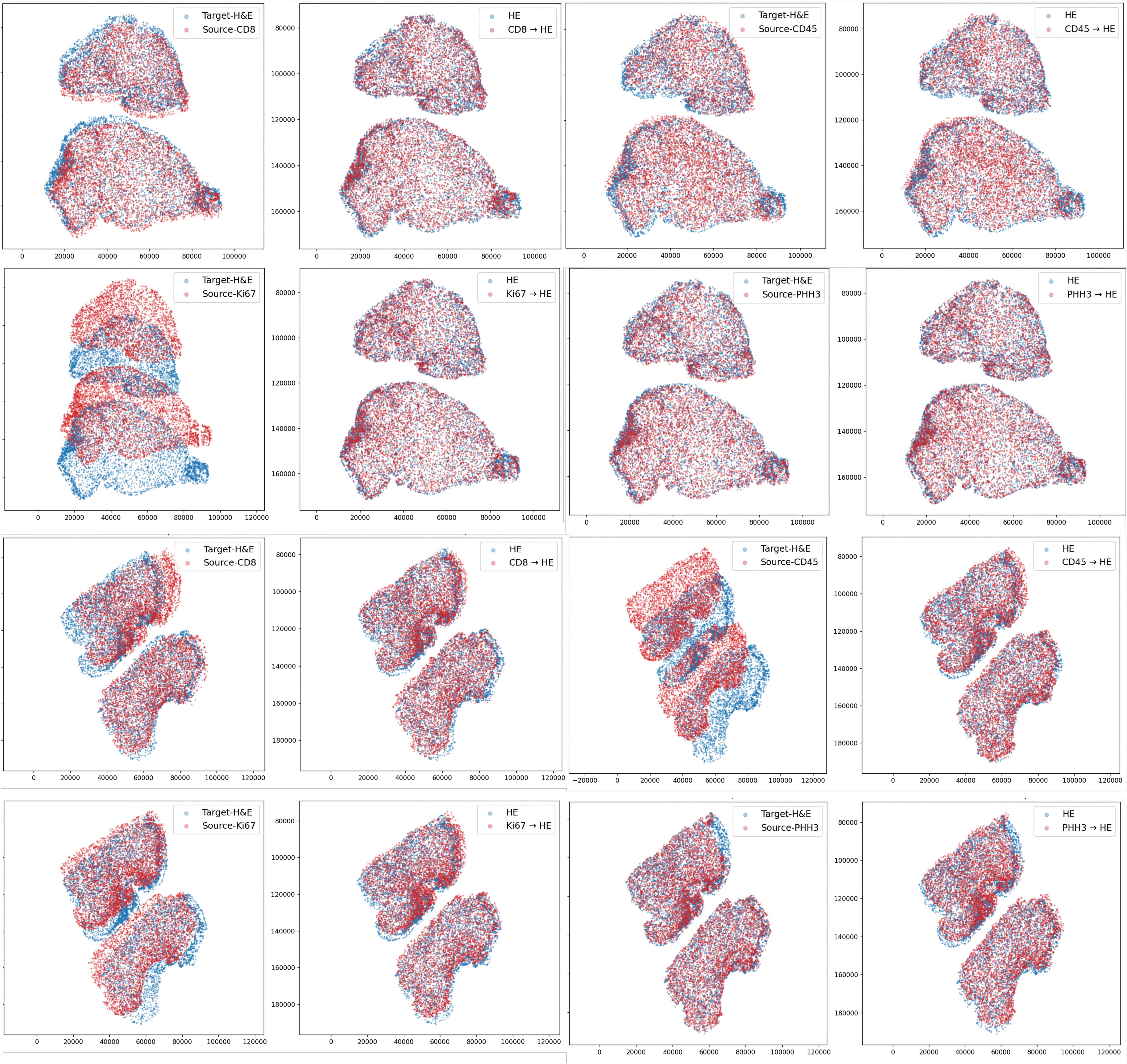}
    \captionsetup{width=\textwidth}
    \caption{Shape-aware registration of raw point sets without coarse alignment. Each panel shows pairwise registration results between nuclei-derived point clouds, with red and blue denoting the target and source domains, respectively. The method achieves accurate cross-modal alignment across multiple tissue sections and immunostaining conditions (e.g., CD8, KI67, CD45, and PHH3), preserving local structural correspondence under significant spatial and staining variability.}
    \label{fig:pointset_results}
\end{figure}
\begin{figure}[H]
    \centering
    \includegraphics[width=1\textwidth]{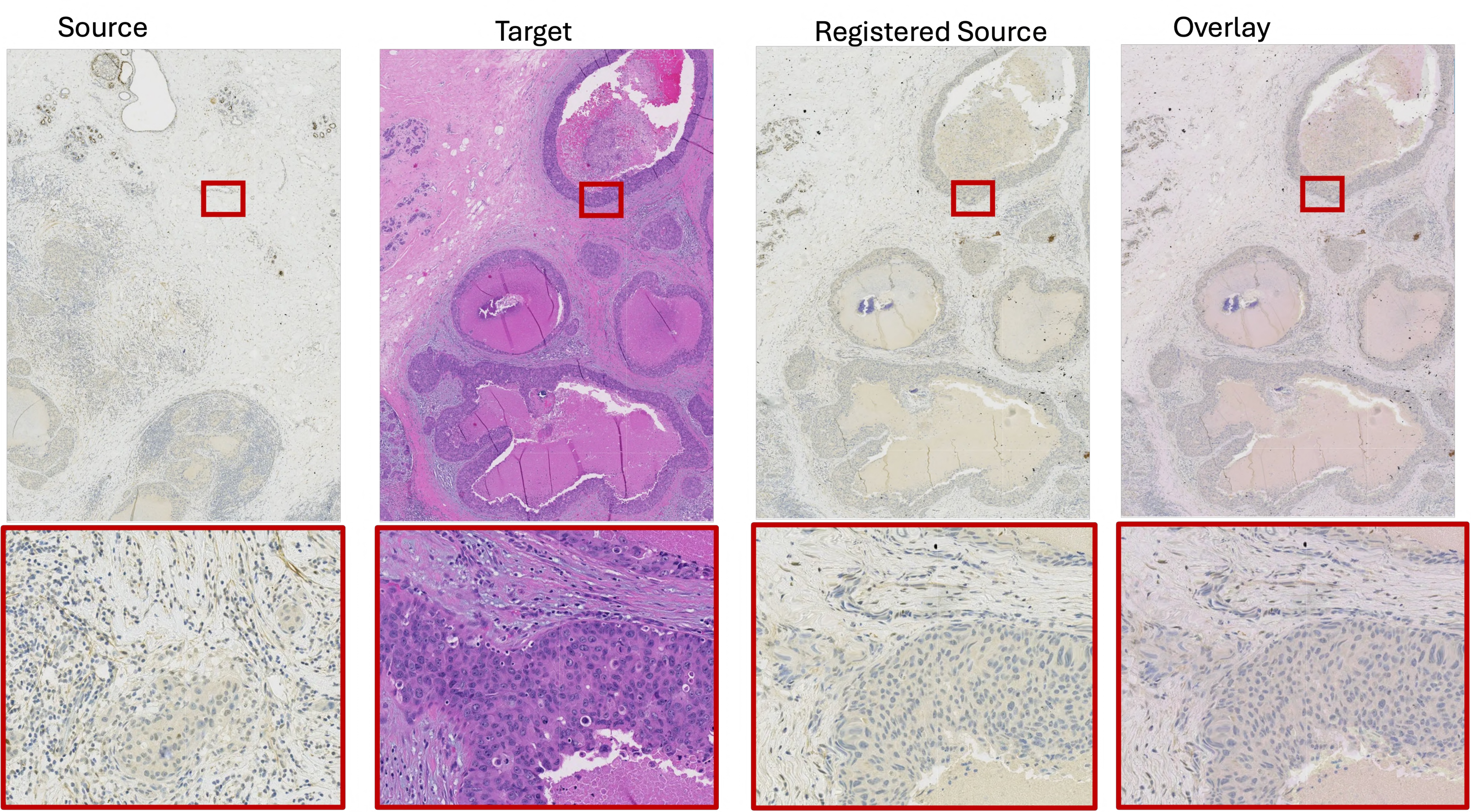}
    \captionsetup{width=\textwidth}
    \caption{CORE registration visualization in presence of tissue folds artifacts. Left (Source) ER WSI region from ACROBAT dataset, (Target) H\&E stain, Right  Coarse Registered Source and Overlay. Zoomed Region visualized in Red Square blocks. }
    \label{fig:acrobat_reg_artifacts}
\end{figure}
\begin{figure}[H]
    \centering
    \includegraphics[width=1\textwidth]{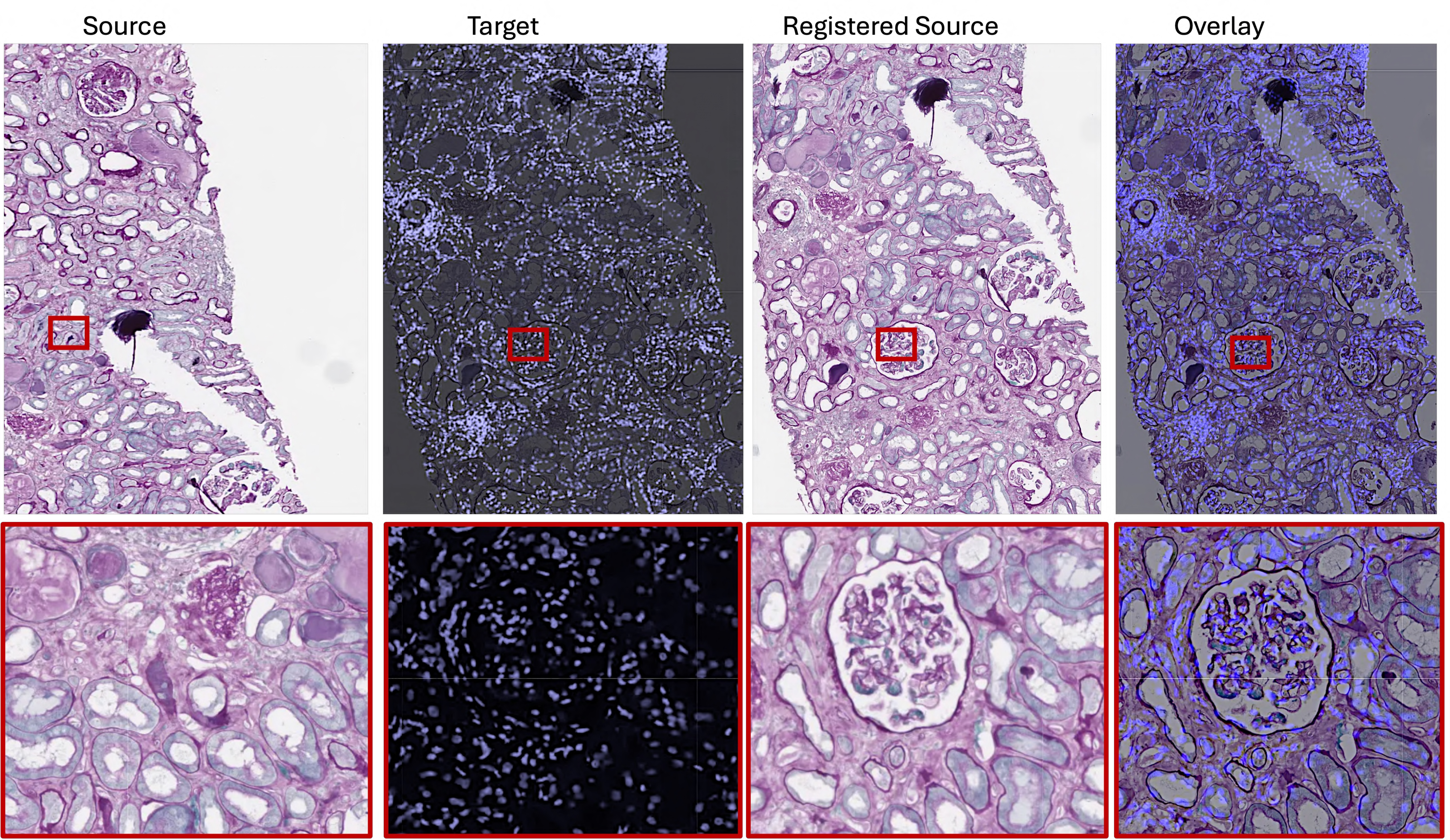}
    \captionsetup{width=\textwidth}
    \caption{CORE registration visualization in presence of missing tissue artifacts. Left (Source) PAS WSI region from REACTIVAS dataset, (Target) mIF stain, Right  Coarse Registered Source and Overlay. Zoomed Region visualized in Red Square blocks.}
    \label{fig:reg_mIF}
\end{figure}
\begin{figure}[H]
    \centering
    \includegraphics[width=1\textwidth]{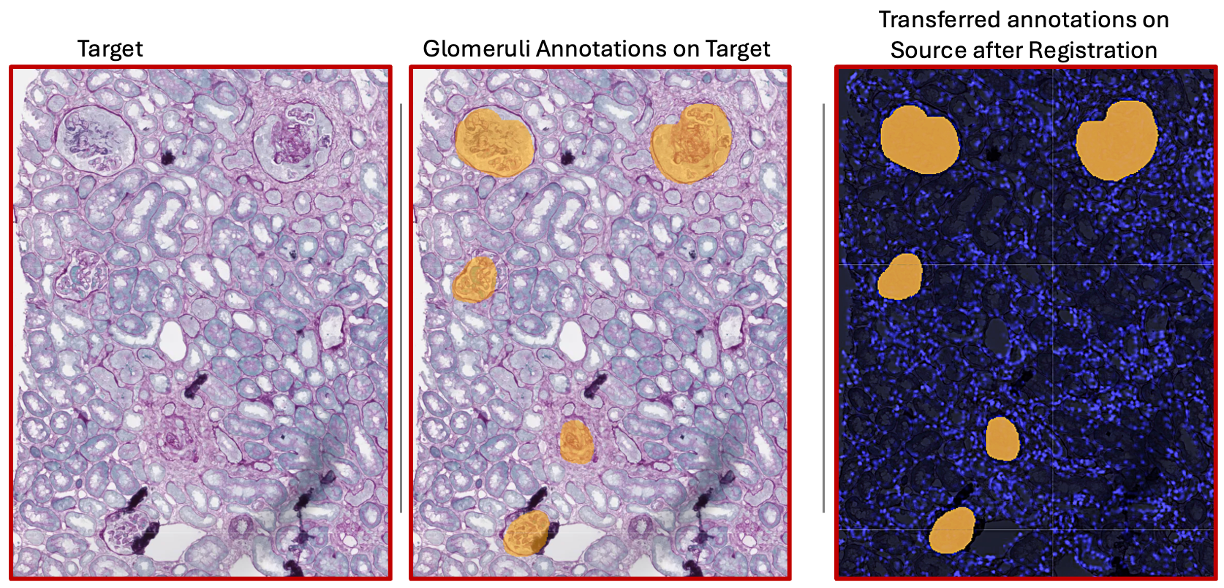}
    \captionsetup{width=\textwidth}
    \caption{ Visualization of glomeruli annotation transfer across multi-stained kidney tissue sections using CORE registration. The left panel shows the target tissue section, the middle panel shows manually delineated glomeruli on the target image, and the right panel shows the corresponding annotations transferred to the source section after registration, demonstrating accurate spatial correspondence between tissue sections. }
    \label{fig:annotation_transfer}
\end{figure}
\end{document}